\documentclass[10pt,journal,comsoc]{IEEEtran}

\ifCLASSOPTIONcompsoc
  \usepackage{cite}
\else
  \usepackage{cite}
\fi

\ifCLASSINFOpdf
  \usepackage[pdftex]{graphicx}
\else
\fi

\usepackage{amsmath}
\usepackage{amsthm}

\usepackage{algorithm}
\usepackage{algorithmicx}
\usepackage{algpseudocode}

\algnewcommand\algorithmicforeach{\textbf{foreach}}
\algdef{S}[FOR]{ForEach}[1]{\algorithmicforeach\ #1\ \algorithmicdo}

\algnewcommand\Or{\textbf{or} }
\algnewcommand\AAA{\textbf{and} }
\algnewcommand\Continue{\textbf{continue}}
\newtheorem{theorem}{\bf Theorem}

\ifCLASSOPTIONcompsoc
\else
\fi
\usepackage{subcaption}

\usepackage{listings}
\usepackage{enumitem}

\usepackage{textcomp}
\usepackage{xcolor}
\usepackage{amsmath}
\usepackage{caption}
\usepackage{booktabs}
\usepackage{bbm}
\usepackage{mathtools}
\usepackage{color}
\usepackage{dsfont}
\usepackage{amssymb} %

\def\BibTeX{{\rm B\kern-.05em{\sc i\kern-.025em b}\kern-.08em
		T\kern-.1667em\lower.7ex\hbox{E}\kern-.125emX}}

\newcommand{\para}[1]{\noindent {\bf #1}}
\newcommand{\ignore}[1]{}

\begin{document}
\title{From Simulation to Reality: Practical Deep Reinforcement Learning-based Link Adaptation for Cellular Networks}

\author{
	\IEEEauthorblockN{
		Lizhao You\IEEEauthorrefmark{2},
		Nanqing Zhou\IEEEauthorrefmark{2}\IEEEauthorrefmark{3},
		Guanglong Pang\IEEEauthorrefmark{2},
		Jiajie Huang\IEEEauthorrefmark{2},
		Yulin Shao\IEEEauthorrefmark{4}, 
		Liqun Fu\IEEEauthorrefmark{2}\textsuperscript{*}\\
	}
	\IEEEauthorblockA{
		\IEEEauthorrefmark{2}School of Informatics, Xiamen University, China \quad \\
		\IEEEauthorrefmark{3}ArrayComm Wireless Technologies Co., Ltd., China \\
		\IEEEauthorrefmark{4}Department of Electrical and Computer Engineering, The University of Hong Kong, China\\
	}
	\thanks{Liqun Fu is the corresponding author (email: liqun@xmu.edu.cn).}
}

\IEEEtitleabstractindextext{
\begin{abstract}
Link Adaptation (LA) that dynamically adjusts the Modulation and Coding Schemes (MCS) to accommodate time-varying channels is crucial and challenging in cellular networks. Deep reinforcement learning (DRL)-based LA that learns to make decision through the interaction with the environment is a promising approach to improve throughput. However, existing DRL-based LA algorithms are typically evaluated in simplified simulation environments, neglecting practical issues such as ACK/NACK feedback delay, retransmission and parallel hybrid automatic repeat request (HARQ). Moreover, these algorithms overlook the impact of DRL execution latency, which can significantly degrade system performance.
To address these challenges, we propose Decoupling-DQN (DC-DQN), a new DRL framework that separates traditional DRL's coupled training and inference processes into two modules based on Deep Q Networks (DQN): a real-time inference module and an out-of-decision-loop training module. Based on this framework, we introduce a novel DRL-based LA algorithm, DC-DQN-LA. The algorithm incorporates practical considerations by designing state, action, and reward functions that account for feedback delays, parallel HARQ, and retransmissions.
We implemented a prototype using USRP software-defined radios and srsRAN software. Experimental results demonstrate that DC-DQN-LA improves throughput by 40\% to 70\%  in mobile scenario compared with baseline LA algorithms, while maintaining comparable block error rates, and can quickly adapt to environment changes in mobile-to-static scenario. These results highlight the efficiency and practicality of the proposed DRL-based LA algorithm.
\end{abstract}

}

\maketitle

\IEEEdisplaynontitleabstractindextext

\IEEEpeerreviewmaketitle

\section{Introduction}\label{sec:intro}

Link adaptation (LA) is a critical technology for enhancing the efficiency of cellular networks by dynamically adjusting the modulation and coding schemes (MCS) for each transmission to accommodate time-varying wireless channel conditions. 
Incorrect MCS decisions can result in either packet loss or a reduction in throughput.
Compared to uplink LA, downlink LA poses greater challenges because the base station (BS) cannot directly measure the downlink channel quality and must rely on channel feedback such as Channel Quality Indicator (CQI) reports from user equipment (UE). This paper focuses on tackling the problem of downlink LA.

Inner-loop link adaptation (ILLA) and outer loop link adaptation (OLLA) are two most common LA approaches used in practice. The key idea is to map the reported CQIs (and the ACK/NACK information) to a MCS according to predefined tables and fixed algorithm parameters. However, using predefined tables and fixed parameters restricts their adaptability in different environments \cite{target_bler}. 
Learning-based LA adapts to the environment through the interaction with the environment, and is a promising approach to mitigate the problem in complex environments.
Ref. \cite{bayes} adopts the Bayesian approach to directly model the MCS success probabilities conditioned on the CQI through a suitable probability distribution, and finds the optimal MCS by the distribution.
Refs. \cite{2019drl,ye2023tmc,parsa2022joint} further present deep reinforcement learning(DRL)-based link adaptation  (DRLLA) algorithms in the presence of periodical and outdated CQI feedback, demonstrating throughput improvement compared with existing approaches in simulated environments.

Although these DRLLA algorithms have demonstrated superior performance in simulations, none real-time DRLLA system is reported. There are two severe limitation when applying in practical systems. First, they ignore the computation latency of the adopted DRL technology. The latency is critical to the LA performance in practice: if the channel changes during the time, the outdated decision may lead to severe performance degradation.
Unfortunately, the online DRL algorithm usually includes a training process and an inference process, and these two processes are executed sequentially. The execution time of DRLLA algorithms (including training and inference) can easily exceed the duration of a TTI in the 4G/5G cellular system ranging from 0.0625 ms to 1 ms. According to our simulation results in Section \ref{sec:motivation}, execution latency indeed has a significant impact on the achieved throughput. 

Another problem is that existing DRLLA algorithms do not consider hybrid automatic repeat request (HARQ) in practical cellular systems. In particular, they simply assume the packet is either received (and ACKed) or dropped (and NACKed) in one round. However, in practical systems, packets failing to transmit successfully are re-transmitted until successful transmission. In this case, the received ACKs/NACKs in the first transmission and re-transmissions have different meanings, which cannot be treated equally. Moreover, the cellular systems have multiple parallel HARQ processes due to the feedback delay between the downlink and the uplink, and thus the decision outcomes are interleaved. The designed DRL algorithm should be robust to the outdated feedback and account for the parallel HARQ transmissions and retransmissions.

In this paper, we provide the first practical DRL-based LA algorithm and a real-time cellular system based on the algorithm. 
We first present a new DRL framework based on DQN that separates the DRL training and inference processes for quick response. The new DQN framework is referred to \emph{Decoupling-DQN} (or \emph{DC-DQN}). Then we present a new DC-DQN-based LA algorithm named \emph{DC-DQN-LA}. In particular, DC-DQN-LA adopts a two-level framework comprising: i) a lightweight inference module that is inside the network stack and responds to received CQIs and ACKs/NACKs to achieve fine-grained, real-time adjustment in MCS selection; and ii) a training module that is outside the network stack and regularly selects experiences from the experience pool for parameter tuning. Parameters of the inference module are regularly updated by the training module to adapt to changing link conditions. In this way, we can ensure that the execution latency of the DRL-based LA algorithm does not hurt the system performance.

Furthermore, we carefully craft the DC-DQN-LA algorithm to fit the practical cellular systems with the following designs: 1) we employ a diverse range of feedback data, including CQI, CQI difference, (N)ACK feedback, and historical information, to describe the complete channel state; %
2) we track the number of retransmissions for each transmission block (TB) and adjust the impact of retransmissions on the reward function by scaling it accordingly; and 3) we meticulously synchronize feedback on actions and delays in accordance with fixed delay regulations for accurate reward calculation and state updates, and package this information into training experiences, making the designed algorithm robust to real-world system delays.

We have implemented a complete DC-DQN-LA prototype on the USRP software-defined radio (SDR) devices using the srsRAN 4G LTE software \cite{srslte}. The inference module is integrated into the data link layer of srsRAN, and the training module is implemented outside srsRAN. %
Based on this prototype, we conduct multiple real-world experiments in both static and mobile indoor environments and collect the channel SNR traces. We also implement a discrete-event network simulator to emulate the protocol in cellular networks based on the measured channel SNR. Experimental results show that DC-DQN-LA outperforms the baseline algorithms (BayesLA \cite{bayes} and OLLA \cite{hsdpa}) by 40\% and 70\%, respectively in terms of throughput in mobile scenario. In static scenario, DC-DQN-LA achieves slightly better performance than the baseline algorithms. DC-DQN-LA can quickly adapt to environment changes in mobile-to-static scenario.
These results demonstrate that DC-DQN-LA consistently achieves high performance with fast and stable inference time and low control overhead, exhibiting good adaptability to various real-world environments.

Overall, this paper makes the following contributions:
\begin{itemize}
    \item We present an online DRL implementation framework that decouples training and inference to enable real-time decision in practical communication systems.
    \item We present the first practical DRL-based LA algorithm that accounts for issues in real cellular networks such as delays, retransmissions and parallel HARQ processes.
    \item We implement a prototype on software-defined radio devices, and demonstrate the throughput improvement over the baseline LA algorithms through experiments.
\end{itemize}

\section{Background and Motivation} \label{sec:motivation}
\subsection{Link Adaptation in Cellular Networks}

Cellular networks utilize the link adaptation technology to adjust the transmission parameters to accommodate the time-varying wireless channel. In particular, we focus on the downlink data transmission from BS to UE, and how BS selects the MCS for each downlink transmission. 
We consider a fully buffered scenario, where BS has sufficient data awaiting transmission.

\begin{figure}[t!] %
\centerline{\includegraphics[width=0.48\textwidth]{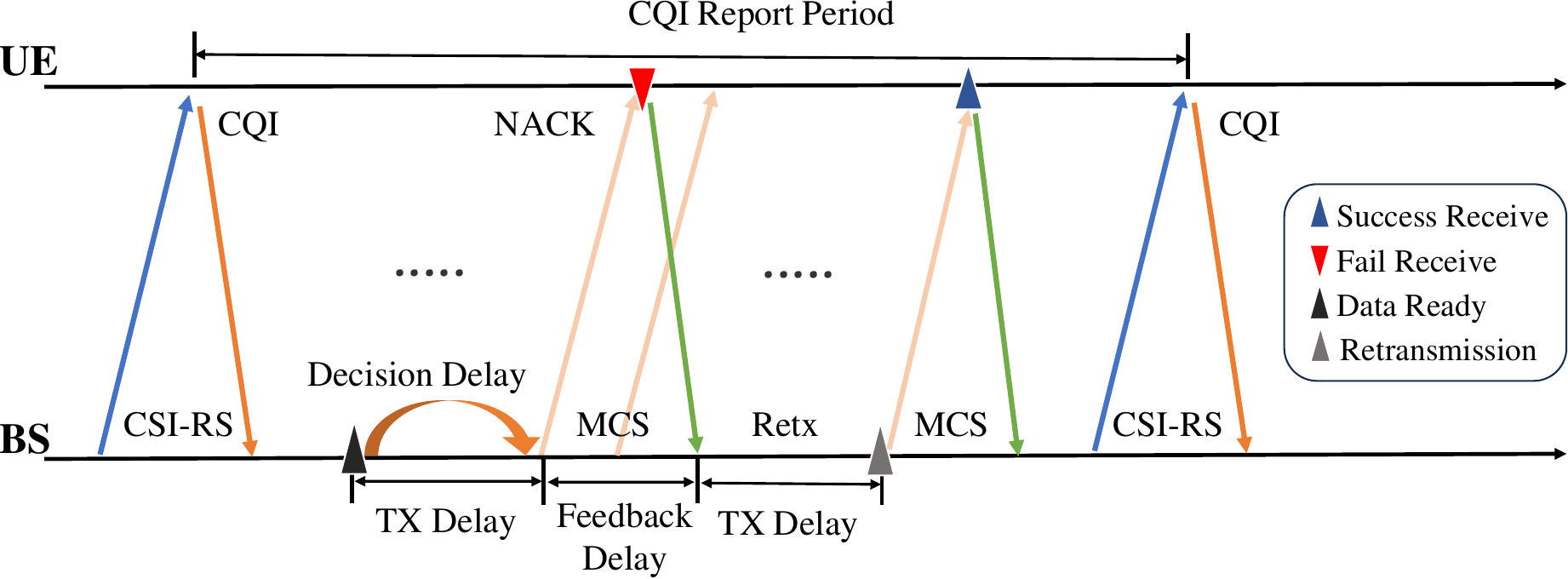}}  %
\caption{Illustration of the link adaptation interaction process of a BS and an UE.}
\label{fig interaction}
\end{figure}

Fig.~\ref{fig interaction} shows a simplified interaction process of a BS and an UE for link adaptation. BS sends a Channel State Information - Reference Signal (CSI-RS) periodically or aperiodically. UE measures the CSI-RS to compute the average signal-to-noise ratio (SNR), and maps the measured SNR to a 4-bit length Channel Quality Indicator (CQI) according to a pre-defined mapping table. 
Then, UE encapsulates these information into a CQI report and transmits it back to BS. 
To reduce the protocol overhead, UE does not feedback CQI instantly when the CSI-RS is received. Instead, there is a parameter named \emph{CQI reporting period} defining the time between two CQI feedback. 
The reported CQI is the average CQI computed over the reporting period. %

The BS schedules UEs to transmit data during each transmission time interval (TTI). In a typical implementation, the BS employs a resource allocation algorithm, such as the proportional fair scheduler, to determine which UEs are scheduled and to allocate the corresponding resource blocks (RBs) to them. Simultaneously, the BS selects the modulation and coding scheme (MCS) for each scheduled UE based on the algorithm described in Section \ref{sec:motivation}. Since UEs are scheduled on orthogonal RBs, their transmissions can be treated as independent processes after RB allocation. \emph{Thus, for simplicity, we focus on a single UE to analyze the link adaptation problem.}

The selected MCS defines a channel coding scheme and modulation scheme used to carry data in each resource element (i.e., subcarrier). Typically, there is a total of $M=28$ MCS schemes available indexed from 0 to 27. After determining the MCS index, the information bits for transmission are encoded into a variable-sized TB. The size of the TB is determined by the number of allocated RB and the adopted MCS scheme. BS conveys the configured parameters in downlink control information (DCI) and transmits it the UE. Finally, BS schedules the downlink transmission at TTI $t$ using $MCS_t$, and the scheduled UE follows the configuration in DCI to decode the information bits. %

The link adaptation algorithm may choose an unsuitable MCS, leading to a decoding error in UE. To achieve reliable communication, cellular networks employ the HARQ technology. Upon decoding the information bits, the UE performs a cyclic redundancy check on the recovered data bits to determine whether the received TB is correct. Then, the UE notifies the BS of transmission status through a 1-bit feedback: ACK (success) or NACK (failure). In the event of transmission failure (i.e., NACK), the corresponding data packets are added to the retransmission queue and wait for retransmission. Retransmission typically reuses the previous MCS, i.e., $MCS_{t+d}=MCS_{t}$ where $t+d$ is the retransmission TTI corresponding to the transmission at the $t$-th TTI. 
Cellular networks supports HARQ based on chase combining (CC) and incremental redundancy (IR).
Each transmission is marked with a redundancy version (rvidx), indicating whether the currently transmitted TB is a new transmission or a retransmission. The system allows at most three retransmissions. 

In practical cellular systems, there exist many delays mandated by system specifications due to the computation constraints. The \emph{TX delay} is defined to be the delay between the time when the UE's data is available and the time when the UE's data is transmitted. The TX delay accounts for the link-layer scheduling latency and the data modulation and coding latency. The link adaptation algorithm also costs a certain amount of computation time. The MCS computation time is referred to as \emph{decision delay}.
An ideal link adaptation algorithm can always make timely decisions before the TX delay. However, if the execution time of the link adaptation algorithm is longer than the TX delay, it may lead to severe performance degradation and even interrupt the connection between the BS and the UE. Hence, in cases where the link adaptation algorithm is unable to make a decision within the defined delay, previous MCS is used to ensure connections.

\subsection{Limitation of Existing Link Adaptation Algorithms} \label{sec:motivation}

There exist some link adaptation algorithms.

\textbf{Inner-loop link adaptation (ILLA).}
In the traditional ILLA algorithm, BS uses the reported CQI to compute MCS directly. However, the algorithm is inaccurate due to the following reasons. First, there exists a \emph{feedback delay} between the downlink transmission and uplink transmission. Thus, the reported CQI only represents the channel at the moment when the CSI-RS is measured, and the channel at the current moment may change.
Second, there exists quantization errors in mapping SNR to CQI at the UE side and in mapping CQI to MCS at the BS side, since the reported CQI has only 4 bits. Hence, relying solely on CQI to determine MCS may be inaccurate.

\textbf{Outer-loop link adaptation (OLLA).}
OLLA is proposed to alleviate the problem. OLLA corrects the MCS decision by adding a SNR offset when mapping the CQI to the MCS at the BS side. In particular, OLLA first maps the received CQI to a SNR, adds a SNR offset to the base SNR, and then maps the offseted SNR to MCS. The SNR offset is adjusted based on the ACK/NACK feedback information. The initial SNR offset is set to 0. When an ACK feedback is received, the SNR offset is added by a step up size $\Delta_{up}$. When an NACK feedback is received, the SNR offset is reduced by a step down size $\Delta_{down}$. These two parameters are usually set according to the following formula
$$\Delta_{down} = \frac{\Delta_{up}}{\frac{1}{BLER}-1},$$
where $BLER$ is the target block error rate to balance the packet loss and the overall throughput. $BLER$ is usually set to 0.1 in many cellular systems.

OLLA relies on the ACK/NACK feedback to adjust SNR offset, and can compensate the inaccurate SNR mapping in some cases. However, the feedback delay makes the ACK/NACK signal outdated, and may lead to a wrong decision. Moreover, the predefined BLER target limits throughput performance since the optimal BLER for maximizing throughput can vary between 10\% and 30\% in different environments \cite{target_bler}. The fixed step size also restricts OLLA's adaptability. A small step size makes OLLA unable to adapt quickly to channel variations. In contrast, although offering rapid convergence, a large step size can lead to significant fluctuations around the optimal value \cite{bayes}.

\begin{figure}[t!]
\centerline{\includegraphics[width=0.45\textwidth]{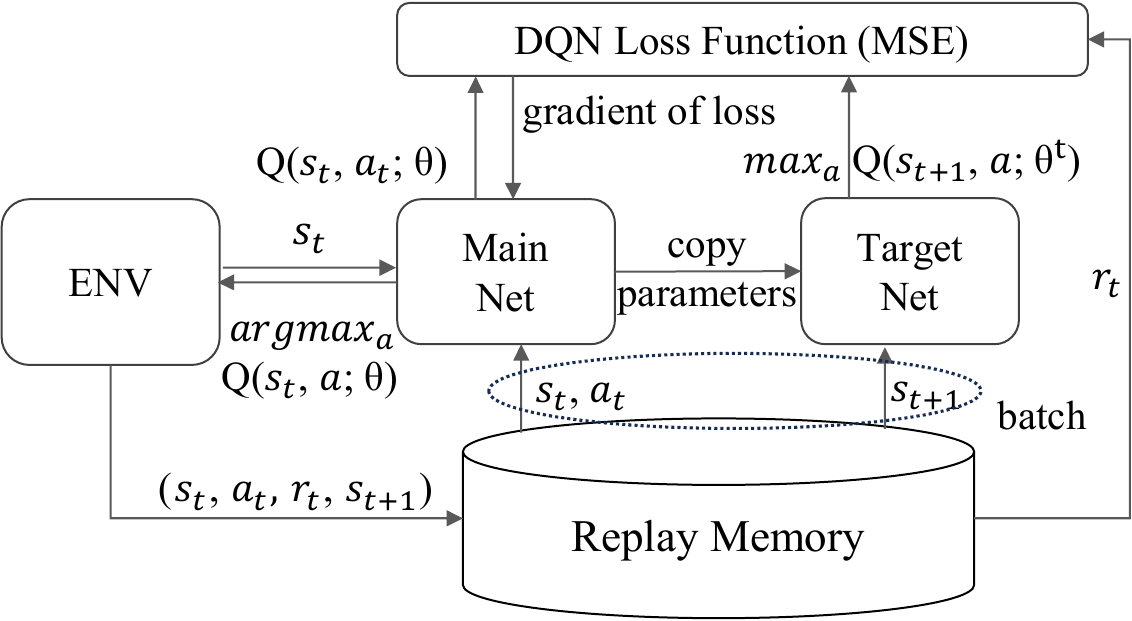}}
\caption{The architecture of the traditional DQN.}
\label{fig framework}
\end{figure}

\textbf{DRL-based link adaptation (DRLLA).}
DRLLA can learn to adapt to different environments, and is a promising approach to optimize MCS selection in complex and time-varying environments. 
In particular, refs. \cite{2019drl,ye2023tmc,parsa2022joint} have demonstrated higher throughput and lower BLER than the OLLA algorithm.

However, all these algorithms are evaluated in simple cellular network environments without considering the practical factors such as the ACK/NACK feedback delay and the HARQ retransmission process. Even worse, these algorithms ignore the computation latency of the adopted DRL technology that tightly couples the training and inference processes.
In scenarios where online training deployment is required, this coupling can lead to delays in decision-making, which are unacceptable in practical systems and may result in significant performance degradation between BS and UE. 

DQN is the most commonly-used DRL algorithm in the literature. Fig.~\ref{fig framework} shows the typical architecture of DQN. DQN employs a neural network to learn the Q-value function. This function maps states and actions to Q-values, representing the expected reward in a specific state by taking that action. 
In DQN, there are two neural networks: the main network, which continuously learns and updates, and the target neural network, which periodically copies the main neural network parameters and calculates Q-Target. The target network ensures stability in the Q-Target value over a short period, preventing the training target from constantly shifting. %

In traditional DQN, inference and training follow a sequential flow, where inference cannot be performed during training, leading to potential delays in decision-making. 
Experimental results in Section \ref{sec:eval:results} show that when the DRL agent selects MCS for a single UE per TTI, the average inference time is approximately 0.5 ms, while the average training time exceeds 5 ms. Considering that the duration of the typical TTI in the 4G and 5G cellular systems ranges from 0.0625 ms to 1 ms\cite{dahlman20134g,5g}, multiple TTIs can occur between receiving state information and executing MCS selection. 
The delay may lead to severe performance degradation, and even communication failures.

We perform a simulation to evaluate the impact of the decision delay on the system throughput. We implement the state-of-the-art (SOTA) DRLLA algorithm proposed in \cite{ye2023tmc}, and use its default and simple environment setup in \cite{ye2023tmc} for evaluation.
We assume the full buffer scenario, the CQI feedback delay 4 TTI, and the CQI reporting period 100 TTI. There is no ACK/NACK feedback delay, and no retransmissions. We assume there exists a decision delay, and the inferred MCS is immediately applied after the decision delay. If the new MCS is not ready, we reuse the previous MCS for the current transmissions.

\begin{figure}[t!]
\centerline{\includegraphics[width=0.49\textwidth]{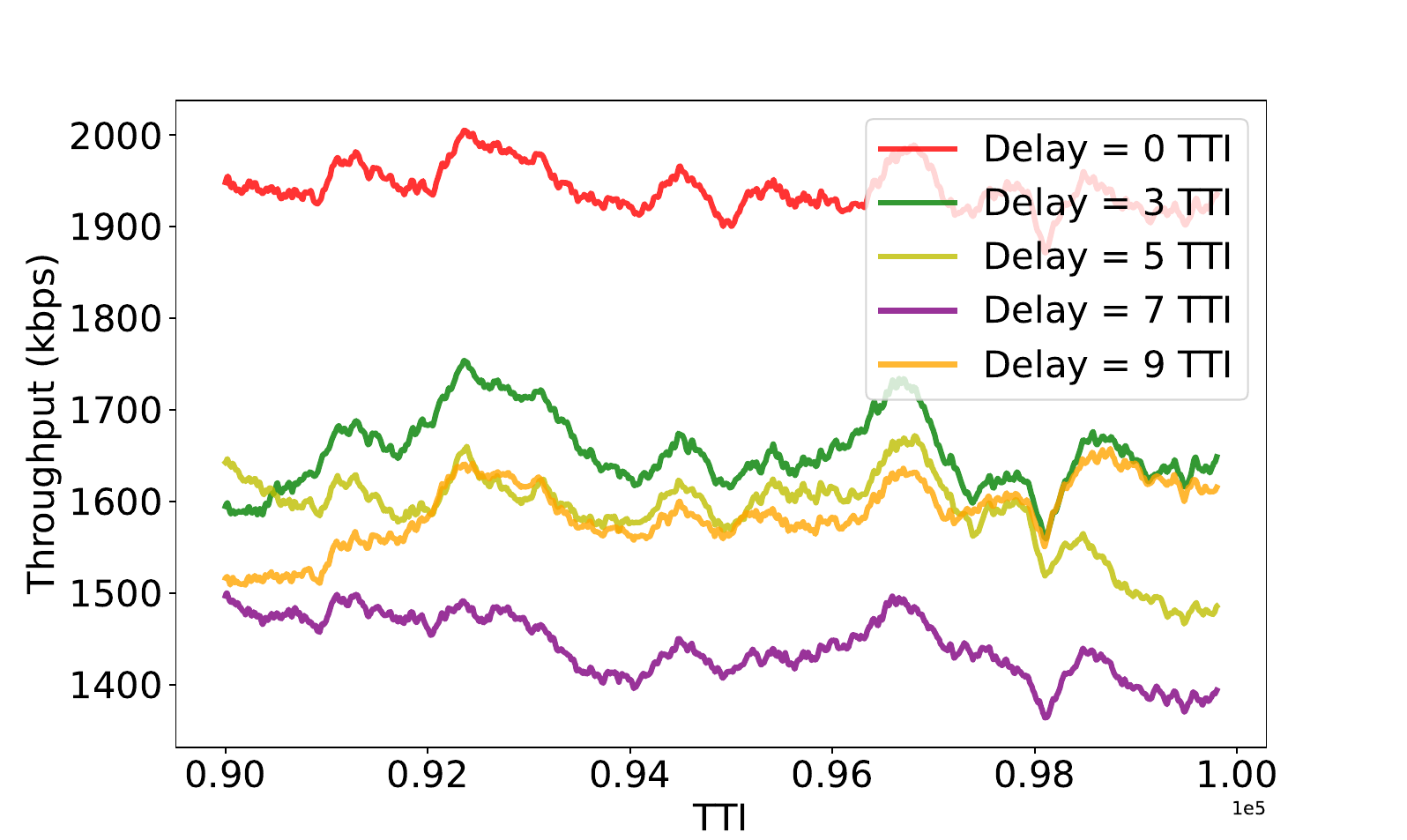}}
\caption{Impact of decision delay on the throughput of the state-of-the-art DRLLA.}
\label{fig dec delay}
\end{figure}

Fig.~\ref{fig dec delay} shows the throughput results with varying decision delays, where the X-axis shows the TTI index and the Y-axis shows the average throughput over a window of 2000 TTIs. We can find that the decision delay has a strong impact on the achieved throughput: the larger the decision delay, the lower the achieved throughput. The simulation results imply that the decision delay needs to be considered when applying the DRLLA algorithm in the simplified cellular networks, not to mention the practical cellular networks.

\section{New DRL Implementation Framework} \label{sec:overview}

To address the decision latency issue, we propose decoupling DRL, a new DRL implementation framework. We first define the new framework using DQN as an example, and then show its convergence.

\subsection{New Framework}
We show the new framework using DQN as the underlying DRL algorithm. The framework is compatible with other DRL algorithms such as Proximal Policy Optimization (PPO), Asynchronous Advantage Actor-Critic (A3C).

\begin{figure}[t!]
\centerline{\includegraphics[width=0.49\textwidth]{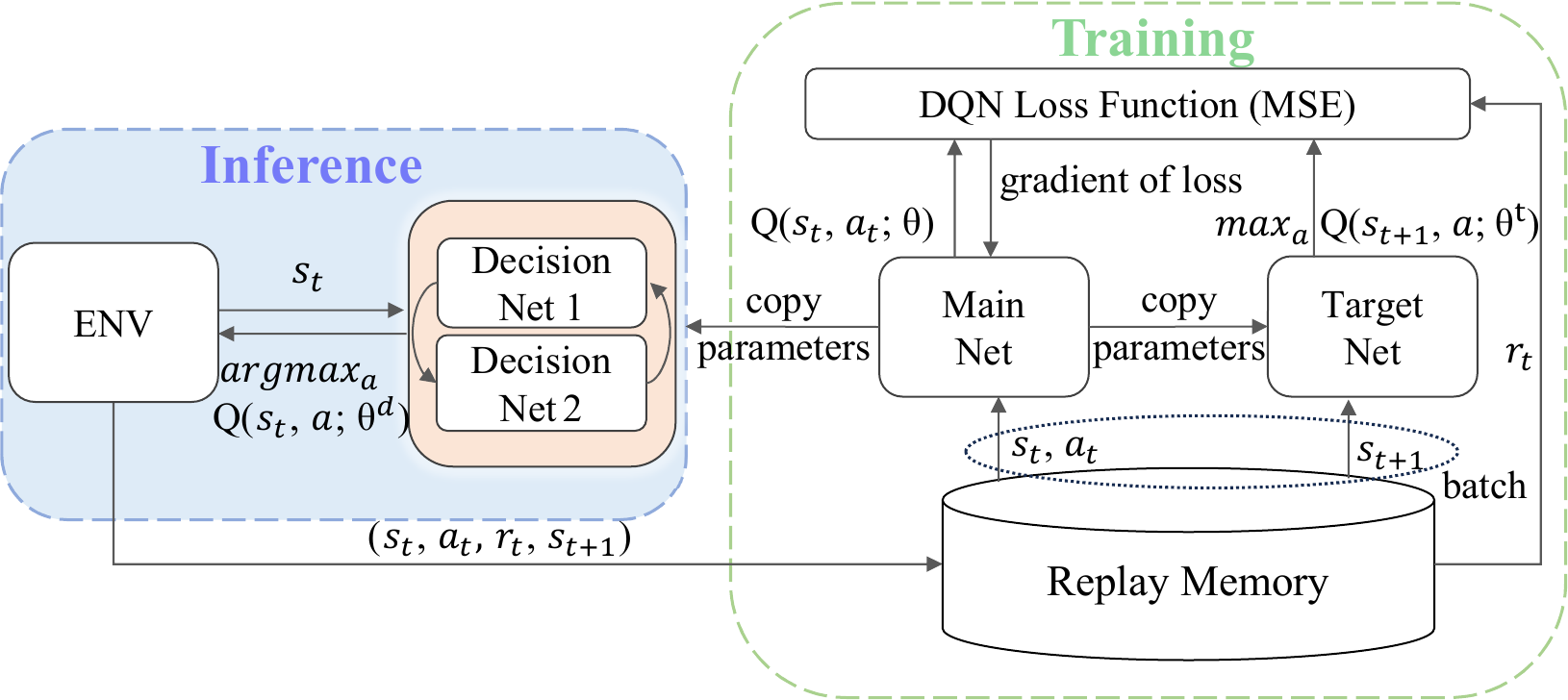}}
\caption{The architecture of the proposed DC-DQN.}
\label{fig framework1}
\end{figure}

Fig.~\ref{fig framework1} shows the architecture of the proposed decoupling DQN (DC-DQN). Compared with the traditional DQN, DC-DQN employs an independent decision-making module for action selection based on the current environmental state. This module generates Q-value estimates for each action, selects the action with the highest Q-value as the output, and regularly updates its parameters from the main network to maintain environmental adaptability.

Considering the time required for the module to receive and update parameters, during which inference is not possible, we design the decision-making module with two decision-making networks, namely decision net1 and decision net2. When one network (net1) makes a decision, the other network (net2) receives main network parameters. Upon parameter reception and detection of a completed inference, the networks exchange their responsibilities, with net1 receiving parameters and net2 performing inference and outputting action. This cyclic exchange of decision-making networks realizes our goal of fully decoupling reasoning and parameter adjustment, effectively controlling the model's inference delay.

This architecture completely decouples the time-consuming training process from the fast inference process, enabling continuous reasoning and leveraging DRL for performance gains. It allows flexible training frequency adjustment, reduces computational resource usage, and enhances environmental adaptability. By executing training and inference asynchronously, DC-DQN optimizes resource efficiency while maintaining performance, ensuring practical deployment efficiency.

In the context of link adaptation, DC-DQN-LA employs hierarchical timing control logic to partition the DRL-based link adaptation algorithm into two distinct subtasks operating at different time scales, with each subtask processed by separate modules. %
The inference module operates at a high frequency to provide fine-grained control, swiftly responding to dynamic channel change. Its primary focus is to make real-time decisions regarding MCS selection based on current link conditions (such as CQI and ACK/NACK) at each TTI. %
The training network module operates at a lower frequency, periodically selecting experiences from the experience pool for training at defined intervals. It systematically adjusts model parameters based on accumulated experiences to effectively adapt to changes in the channel environment. The updated parameters are then copied to the inference module.

\subsection{Convergence Analysis}

We are to show that if the traditional DRL framework can converge, the new framework can also converge. We show the intuition below, and the formal proof is in Appendix A.

\textbf{Traditional DRL.}
The action-value function in DRL is approximated by the main Q-network $Q_{\theta}$ with parameters $\theta$. Let $\theta^{t}$ be the parameters of a target Q-network that is initially identical to the main Q-network, and updated periodically. 
We define the action-value function as the sum of discounted rewards from the current time step $t$ with state $s_t=s$ and action $a_t=a$ until the future, i.e.,

\begin{equation}
Q\left(s,a;\theta\right)=\mathbb{E}\left[\sum_{\tau=0}^\infty\gamma^\tau r_{t+\tau}|{s}_{t}=s,a_{t}=a; \theta \right],
\label{eq converge 1}
\end{equation}
where $r_t$ is the reward received at time step $t.$ We also define
the operator $\mathcal{B}$ as the Bellman operator given as
\begin{equation}
    \mathcal{B}Q(\mathbf{s}_{ t},a_t;\theta)=r_t+\gamma\cdot\max_a Q(\mathbf{s}_{t+1},a;\theta).
\end{equation}
With a batch of samples $\mathcal{L}$ with size $N_{E}$, the loss function $L(\theta)$ is defined as the error between the Bellman reward on the target Q-network and the approximated future reward on the main Q-network, i.e.,
\begin{equation} \label{eq:loss}
    L(\theta)=\frac{1}{N_E}\sum_{t \in \mathcal{L}}\left(r_t+\gamma\max_a Q_{}(s_{t+1},a;\theta^t)-Q(s_{t},a_t;\theta)\right)^2,
\end{equation}
and we use gradients $\nabla_\theta L(\theta)$ to update $\theta$.

\textbf{Decoupling DRL.}
Our decoupling DRL introduces a new decision network with parameters $\theta^{d}$. The parameters $\theta^{d}$ are copied from the parameters of the main Q-network $\theta$ periodically.

Compared with the traditional DRL, the main Q-network still uses \eqref{eq:loss} to update $\theta$. The difference lies in the batch construction, where the batch in decoupling DRL $\mathcal{L}^{dc}$ is constructed based on $\theta^{d}$. In particular, our decoupling DRL chooses the action based on the following formula
\begin{equation} \label{eq:decision-net}
    a_{t}=\max_aQ_{\theta^{d}}(s_{t},a).
\end{equation}

However, compared with the batch construction in traditional DRL $\mathcal{L}$, although the experiences are different, the distribution remains the same since they are still sampled from the environment. Therefore, if there exists an optimal policy (or $\theta$) in the traditional DRL, $\theta$ in the decoupling DRL can still be trained to approach the optimal policy.

\section{DC-DQN-LA Design}
We present a new LA algorithm named DC-DQN-LA based on the new DRL framework, including the definition and the adopted neural network. 

\subsection{MDP Definition} \label{sec:design:mdp}

DRL is usually formulated as a Markov Decision Process (MDP), and is executed in an agent. MCS selection is performed at the BS only, and thus we consider the single-agent setup. Without loss of generality, we assume that training happens in each TTI.
Then, we define the state space, the action space, and the reward function for the $t$-th TTI (round) accordingly.  %

\subsubsection{State Space} %
The state encompasses the environmental observations made by the agent and serves as the foundation for decision-making. The state space, denoted as $State$, encompasses all conceivable states. 
Specifically, the state comprises the following four key pieces of information:
\begin{itemize}
\item CQI $c_{t} \in [0, 15]$. The first information is the most recent CQI received by BS. CQIs, periodically transmitted by UE, provide insights into the current downlink channel quality. Despite potentially outdated and imprecise, the CQI value still offers a description of the present channel condition. Higher CQI values signify favorable channel conditions, allowing for the selection of more aggressive MCS with higher data rate and increased throughput. Conversely, lower CQI values indicate relatively poor channel conditions, necessitating a more conservative MCS selection to avoid packet failures.

\item CQI difference $\delta_{t}$. The second information is the difference between the last two CQIs at the BS, which offers insights into the channel's evolving trend to some extent. Given the dynamic and swiftly altering nature of the communication channel, capturing the overall channel trend from a single moment's channel status is challenging. 
The difference of CQI serves as a guiding principle for the agent's exploration. A positive CQI difference suggests potential improvement in channel conditions, while a negative difference indicates a deteriorating trend, favoring a more conservative MCS option.

\item ACK $ack_{t} \in \{0,1\}$ and MCS selection $m_{t} \in [0, 27]$.  The ACK or NACK signal, along with the selected MCS for a given transmission, constitute crucial information in assessing the link conditions. 
When there are many ACKs, it means that most of the MCS selections are reasonable, and the channel estimation is accurate. To achieve higher throughput, the MCS selection can be appropriately increased. Conversely, an abundance of NACKs indicates an overly optimistic channel estimation. In such cases, promptly lowering the MCS selection is imperative to avert multiple failed transmissions. %
\end{itemize}

Due to the limitations of outdated information and the rapid fluctuations in channel status, relying solely on current information may inadequately capture channel characteristics. Consequently, to accommodate temporal variations in the channel and enhance the agent's channel state estimation, we devised the state comprising a fixed-length historical feature sequence. This sequence is structured into a two-dimensional array. Assume the $i$-th round execution begins at the $t$-th TTI. Let $d_{tx}$ be the TX delay, and $d_{ack}$ be the feedback delay. Then, the sequence is defined as
 \begin{equation}
s_t=[z_{t},\ z_{t-1},\ ...,z_{t-l}],\label{eq_state}
\end{equation}
\begin{equation}
z_{t}=[c_{t},a_{t+d_{tx}+d_{ack}},m_{t+d_{tx}},\delta_{t}],\label{eq_state}
\end{equation}
where $l$ denotes the length of the historical data utilized. In our implementation, we use the same value as in \cite{ye2021multi}. Note that, due to the inherent system delays, the DRL agent cannot construct $z_t$ until TTI $t+d_{tx}+d_{ack}$.

\subsubsection{Action Space} %
We directly define the agent's actions as the MCS chosen by the BS $m_t$. The MCS is represented by an index value ranging from 0 to 27, as stipulated by the protocol, indicating the encoding and modulation strategy employed for the current data downlink transmission.
Each MCS index corresponds to a special physical transmission rate and a TB size ${tb}_{t}$.
Choosing a higher MCS results in a higher data rate and increased risk of transmission failure. The objective of our algorithm is to select the appropriate MCS for each transmission to maximize the throughout.

\subsubsection{Reward Function} %
The design of the reward function stands as a pivotal component within the DRL algorithm, representing the primary objective of the system. It assigns a numerical value to each action, reflecting the the quality of taking this action in a certain state. This function serves as a means to evaluate the agent's performance following each action, thereby providing feedback from the environment. Our goal is to design a reward function that guides the DRL agent to maximize the throughput of the entire communication process.

Considering the retransmission problem within a real communication system, HARQ mechanisms ensure that packets failing to transmit successfully are retransmitted until successful transmission or until the maximum number of retransmissions is reached, at which point they are discarded. In cellular systems, a redundant retransmission approach allows previously failed data transmissions to increase the success rate during subsequent transmissions. Consequently, even with an excessively high MCS, users may eventually accept the data packet after multiple retransmissions. However, this scenario is harmful to guiding the agent accurately, potentially leading to the blind selection of overly high MCS values.

To address this issue, we set the reward for retransmissions in the following way. We count the number of retransmissions of the current TB as ${rtx}_{t}$ . When a transmission is successful, indicated by a returned \emph{ack} of 1 (i.e., ACK), we provide a positive reward. The reward value is calculated as the size of the TB divided by the number of retransmissions. This design aligns with our goal of encouraging the selection of higher code rates to achieve higher throughput while also incentivizing fewer retransmissions. As larger TBs tend to result in fewer retransmissions, the agent receives greater rewards for such selections. Conversely, if the transmission fails, signified by an \emph{ack} of 0 (i.e., NACK), we impose a penalty of $-1 * {rtx}_{t}$ to control the error rate within an acceptable range. This penalty discourages the agent from consistently opting for high MCS values solely to maximize rewards, which could lead to multiple retransmissions and reduced link efficiency. Instead, it prompts the agent to accurately assess the current channel conditions and select an appropriate MCS. Meanwhile, we define $rb$ as the number of RBs occupied by a single UE during the communication process, and normalize the throughput of the UE on each RB by dividing $rb$.
We define the rewards as
\begin{equation} \label{eq:reward}
r_{t}=\begin{cases}{tb}_{t}\ /\ ({rtx}_{t} * rb),&\text{if}\ ack_t=1 \  (ACK),\\-1 * {rtx}_{t}/\ rb,&\text{if}\ ack_t=0\ (NACK). \end{cases}
\end{equation}

\subsubsection{Experience Alignment} \label{sec:mdp:align}

Our model generates actions via the decision-making network and interacts with the environment. Subsequently, it stores the collected experiences in the experience pool and periodically samples them for main network training. Due to the various system delays, state, action and reward at the $t$-th TTI correspond to the observation in different TTIs. Thus, we need to manually align the observation and construct experience.

Assume the index of the current TTI is $t$. Due to the TX delay, the DRL agent can only observe the corresponding action at TTI $t+d_{tx}$. Due to the delay in ACK feedback, the DRL agent can only obtain information on whether the transmission linked with $S_t$ is successful or failed at TTI $t+d_{tx}+d_{ack}$. Then, at TTI $t+d_{tx}+d_{ack}$, the DRL agent constructs the experience $e=[s_t, a_{t+d_{tx}}, r_{t+d_{tx}+d_{ack}}, s_{t+d_{tx}+d_{ack}}]$, and stores $e$ in the experience buffer.

Given the above design, key part of DC-DQN-LA algorithm is given in Alg. \ref{alg2}.

\begin{figure}[t!]
\centerline{\includegraphics[width=0.45\textwidth]{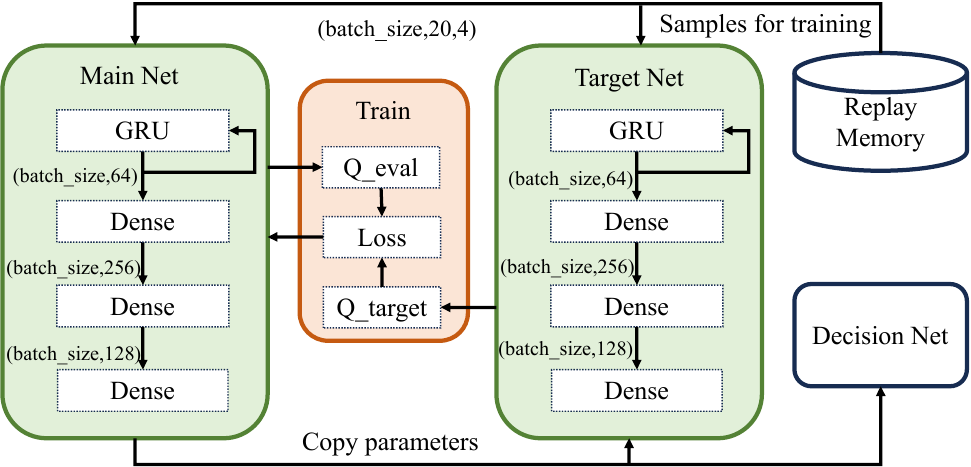}}
\caption{The neural network design of DC-DQN-LA.}
\label{figure archtectrue}
\end{figure}

\subsection{Neural Network Design}
Fig.~\ref{figure archtectrue} illustrates the detailed network architecture of DC-DQN-LA. The DNN model utilized in the DC-DQN-LA implementation comprises four components: an input layer, a GRU layer\cite{cho2014learning}, two Fully Connected (FC) layers, and an output layer. 
The GRU accumulates observations over time, enabling it to capture the underlying temporal characteristics of the channel. As a newer variant of RNNs, GRU excels in inferring complex temporal correlations within input sequences by efficiently storing and reusing historical information. Compared to Long Short-Term Memory (LSTM)\cite{lstm}, another widely used RNN architecture for temporal feature extraction, GRU uses fewer neural parameters, resulting in shorter training times. By embedding GRU into the DQN framework, our algorithm can effectively model the temporal characteristics of long-term historical channel states, enabling dynamic learning of optimal MCS selection strategies in communication systems.

\begin{algorithm}[htb]
    \caption{The DC-DQN-LA Algorithm}
    \label{alg2}
    \begin{algorithmic}[1] %
        \State  Init: main network $\mathcal{Q}$, target networks $\mathcal{Q}^{t}$, decision network $\mathcal{Q}^{d}$ with parameters $\theta$, $\theta^{t}$, $\theta^{d}$, respectively; experience buffer: $buffer$; HARQ retransmission queue: $rtx\_queue$; training interval $T$; network update interval $U$; ACK delay $d_{ack}$; %
        \ForAll{each TTI $t$}
            \State Construct $e$ as in Sec. \ref{sec:mdp:align}, and store $e$ into $buffer$;
            \If{$ack_t$ = 0 (i.e., NACK)}
                \State push $t-d_{ack}$'s TB to $rtx\_queue$;
            \EndIf
            \If{$rtx\_queue$ is not empty}
                \State Pop a TB from $rtx\_queue$;
                \State Use the previous MCS for re-transmission;
                \State Continue;
            \EndIf
            \State Observe $s_{t}$ and compute $a_t$ as in \eqref{eq:decision-net};
            \If{$t$ mod $T$ == 0}
                \State Sample a batch of $N_E$ samples from $buffer$;
                \State Train the main Q-network as in \eqref{eq:loss};
            \EndIf
            \If{$t$ mod $U$ == 0}
                \State  Copy main Q-network's $\theta$ to $\theta^{t}$;
                \State  Copy main Q-network's $\theta$ to $\theta^{d}$;
            \EndIf
        \EndFor
    \end{algorithmic}
\end{algorithm}

\section{Implementation} \label{sec:design:hard}

We implement a BS and a UE using the srsRAN 22.04 software suite \cite{srslte}. In particular, the BS is implemented with srsENB (a full-stack 4G LTE eNodeB) and srsEPC (a light-weight 4G EPC implementation) on a laptop equipped with an Intel Core i7-12700H 2.30 GHz CPU and an NVIDIA RTX 3060 GPU, and the UE is implemented with srsUE on a separate Intel NUC Mini computer. Both BS and UE run on the low-latency Ubuntu 20.04 kernel and connect with an Ettus USRP B210 as a complete transceiver. 

In the following, we elaborate how to incorporate DC-DQN into the existing srsRAN implementation, including the architecture and the interaction flow.

\begin{figure}[t!]
\centerline{\includegraphics[height=4cm]{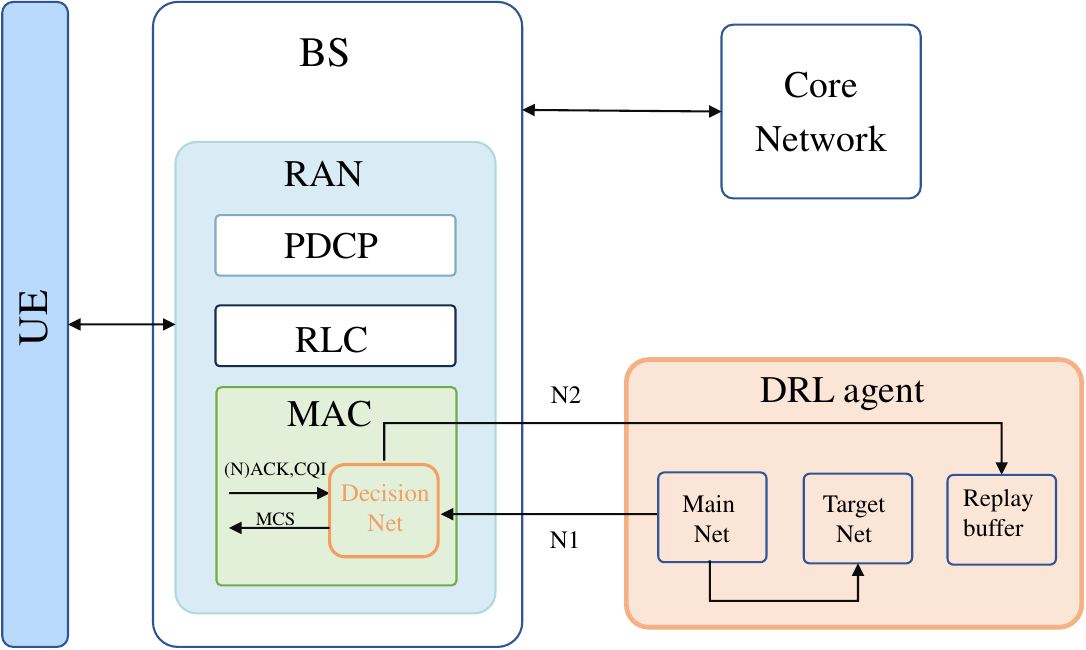}}
\caption{The complete network stack with DC-DQN-LA.}
\label{fig srsran}
\end{figure}

\subsection{Architecture}

Fig.~\ref{fig srsran} shows the complete network stack with DC-DQN-LA in the BS, where the inference module is implemented inside srsRAN and the training module is implemented outside srsRAN.

To realize communication between the DC-DQN-LA inference module and the DC-DQN-LA training module, we have designed a standard-compatible interaction protocol based on TCP, which primarily involves neural network parameter transmission interface $N1$ and experience pool interface $N2$.
We place the training module and experience pool of DC-DQN-LA in the user space, serving as a slow path implementation of the model. The user space allows flexible deployment on either a BS or an edge computing platform. Furthermore, DC-DQN-LA's training module is currently implemented in PyTorch\cite{pytorch}. However, the implementation is not tightly coupled with any specific deep learning framework, and users can optimize neural networks using other frameworks. 
The flexible deployment of training module ensures that it has sufficient computing resources during model training without impacting the normal operation of the communication system.

We position the DC-DQN-LA decision-making module, capable of real-time inference inside srsRAN \cite{srslte}. 
This module is deeply integrated with the MAC layer and implemented using the LibTorch library, a C++ version of Pytorch. %
In this way, the memory load is minimal, and the most of inference time is controlled within 0.5ms, meeting latency requirements in the practical system. The decision-making module tightly integrates with the MAC layer to obtain real-time feedback information such as ACK/NACK and CQI, enabling timely MCS selection for downlink transmission.

Furthermore, the inference module in the network stack serves as a snapshot of the training module deployed in the user space. It regularly receives model parameters from the training module to adapt to the changing channel conditions.
DC-DQN-LA offers full compatibility with existing protocols and incurs lightweight deployment cost. %

\begin{figure}[t!]
\centerline{\includegraphics[height=4.5cm]{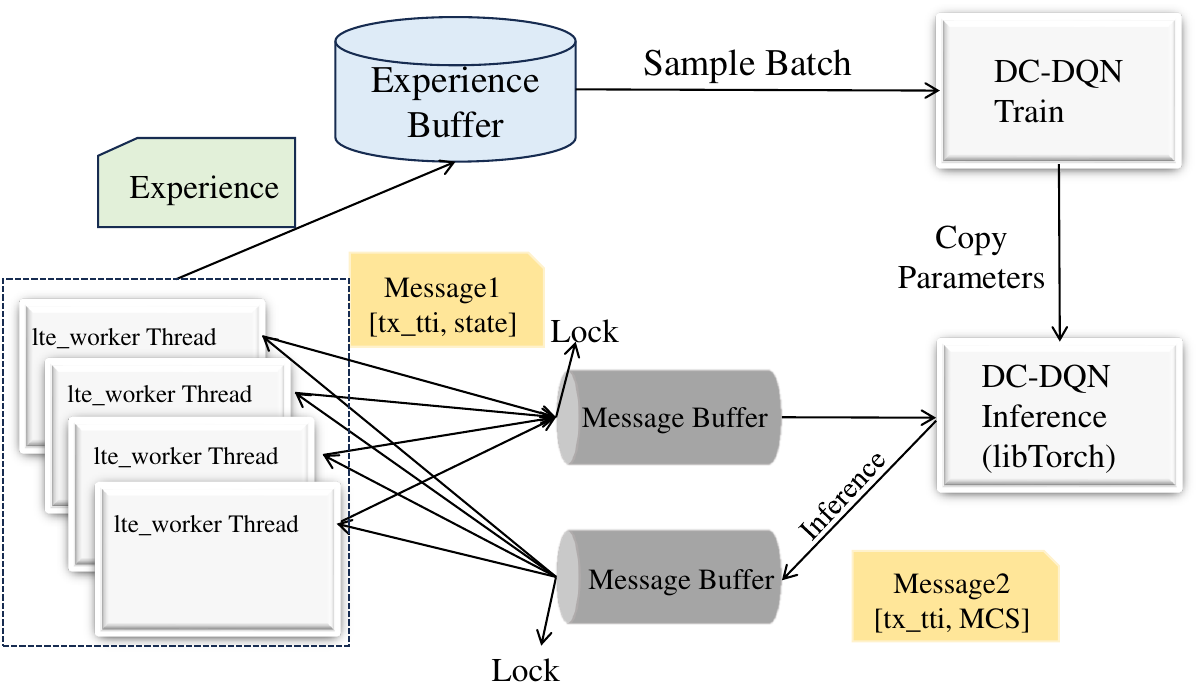}}
\caption{The interaction diagram between DC-DQN-LA and srsRAN.}
\label{fig decison}
\end{figure}

\begin{figure}[t!]
\vspace {-1.5em}
\centerline{\includegraphics[width=0.35\textwidth]{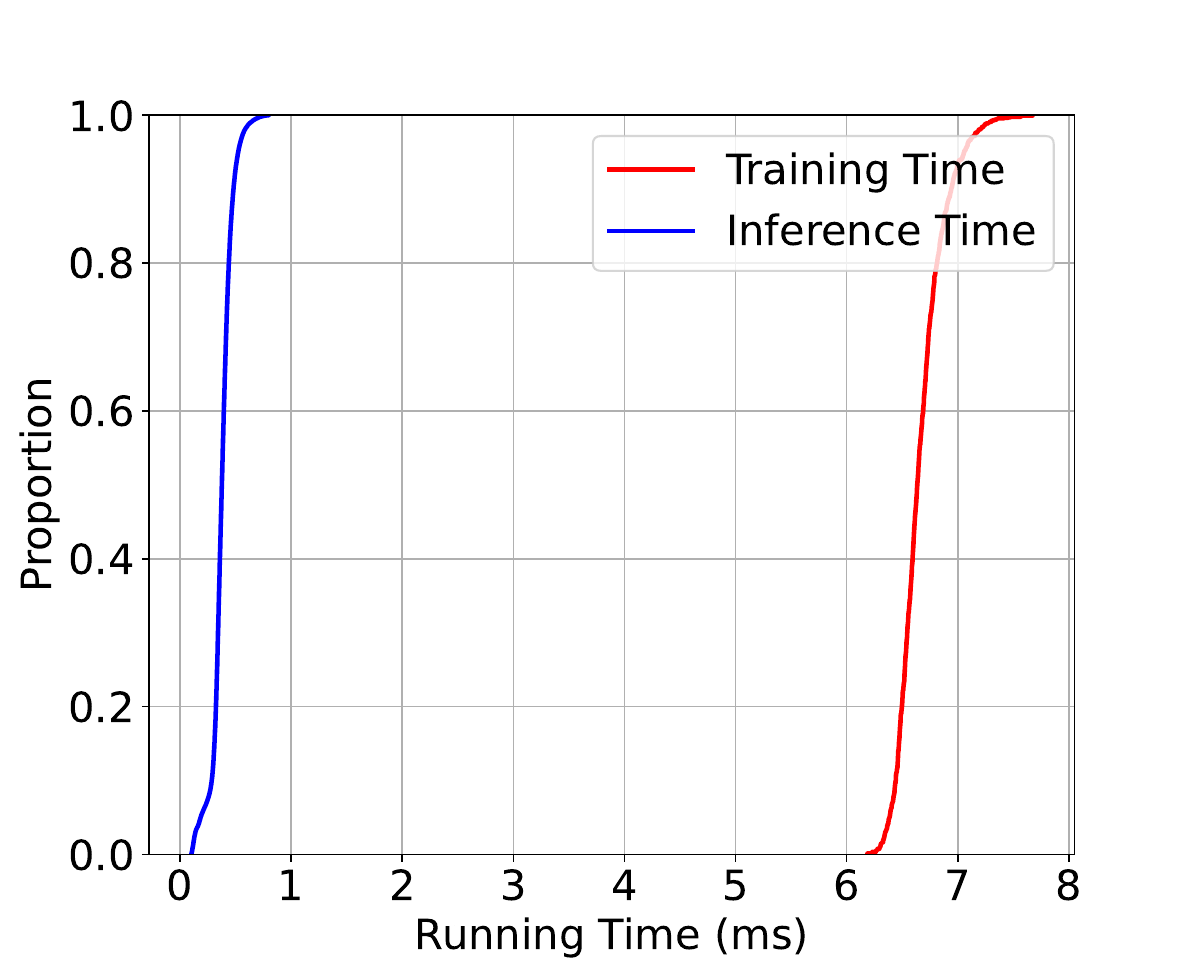}}
\caption{CDF of DC-DQN-LA's inference time and training time in our implementation.}
\label{fig:cdf}
\end{figure}

\subsection{Interaction Flow}
Fig.~\ref{fig decison} shows the interaction flow between DC-DQN-LA and srsRAN. SrsRAN initiates an \emph{lte\_worker} task for each TTI, dispatching these tasks to a thread pool for execution. For the downlink transmission within each task, the TX delay is set to 4 TTIs by default, meaning that the sytem have four TTIs to allocate RB resources, select MCS and encode TBs. Consequently, multiple task threads operate in parallel. To manage delay and ensure sequential decision-making, we allocate the DQN inference module with a dedicated thread. Communication between task threads and the DQN thread is realized through pipelines. %

Specifically, when the \emph{lte\_worker} thread requires an MCS decision, the current status and time slot sequence number are encapsulated into message body 1 and transmitted via the pipeline. The DQN inference module makes decisions based on the received state and time slot order, packaging the MCS selection into a message sent via another pipeline to the task thread. The task thread waits for 0.5ms after issuing a message request while simultaneously monitoring the feedback message pipeline. If no message is received, it indicates an abnormality in the inference module and the previous MCS decision is utilized. This design significantly enhances system robustness, ensuring uninterrupted operation in the event of a DQN inference timeout and effectively enforcing strict latency control.

Our fully-decoupled network architecture for inference and training enables us to meticulously control inference time cost, prioritizing the optimization of the inference network to meet the strict requirements of practical communication systems. The cumulative distribution diagram depicted in Fig.~\ref{fig:cdf} illustrates the distribution of inference time and training time for our algorithm.
It is evident that 90\% of the inference process can be completed within 0.5ms, ensuring that each inference process can be executed within a TTI. Conversely, a training process has the average time required to complete a training iteration is about 6ms.
This clear separation of inference and training processes is crucial. If the inference and training processes are not decoupled, it may happen that inference needs to wait for a training process to be completed, which lead to unpredictable time cost. There would be a risk of failing to meet the system timing requirements. %

\section{Evaluation}

In this section, we evaluate the performance of DC-DQN-LA through experiments conducted on USRPs.
We deploy the algorithm on the SDR-based base station and investigate how DC-DQN-LA can effectively enhance the communication between the UE and BS. 

Besides experiments, we also implement a simplified protocol simulator of the cellular network, and perform controlled simulations to study the impact of system parameters on the algorithm performance. In particular, we adopt the trace-driven simulation approach that uses the collected channel SNR data from our testbed as inputs.

\begin{table}[t!]
    \centering
    \caption{System and DC-DQN-LA parameters} 
    \begin{tabular}{|c|c|c|c|}  
        \hline  
        & & &\\[-6pt]
        \textbf{Parameters}&\textbf{Value}
        &\textbf{Parameters}&\textbf{Value}\\ 
        \hline  
        & & &\\[-6pt]  
        Discount $\gamma$&0.9& \# of CQI levels &16 \\
        \hline  
        & & &\\[-6pt]  
        Learning rate &0.001& \# of MCS schemes &28 \\ %
        \hline  
        & & &\\[-6pt] 
        Training interval $T$ &50 TTIs&ACK delay $d_{ack}$ &8 TTIs \\
        \hline
        & & &\\[-6pt]   
        Update interval $U$ &$T*10$&ENB tx delay $d_{tx}$ &4 TTIs \\
        \hline
        & & &\\[-6pt]    
        State history length $l$ &20 &Bandwidth&10M \\
        \hline
        & & &\\[-6pt]    
        TTI duration&1ms& Mode&FDD \\
        \hline
        & & &\\[-6pt]    
        Max rtx\_num &4&CQI delay &4 TTIs \\
        \hline
        & & &\\[-6pt]   
        Experience buffer size &4096&Batch size& 64\\
        \hline
    \end{tabular}
    \label{table:system-parameters}
\end{table}

\begin{figure}[t!] %
\centerline{\includegraphics[width=0.5\textwidth]{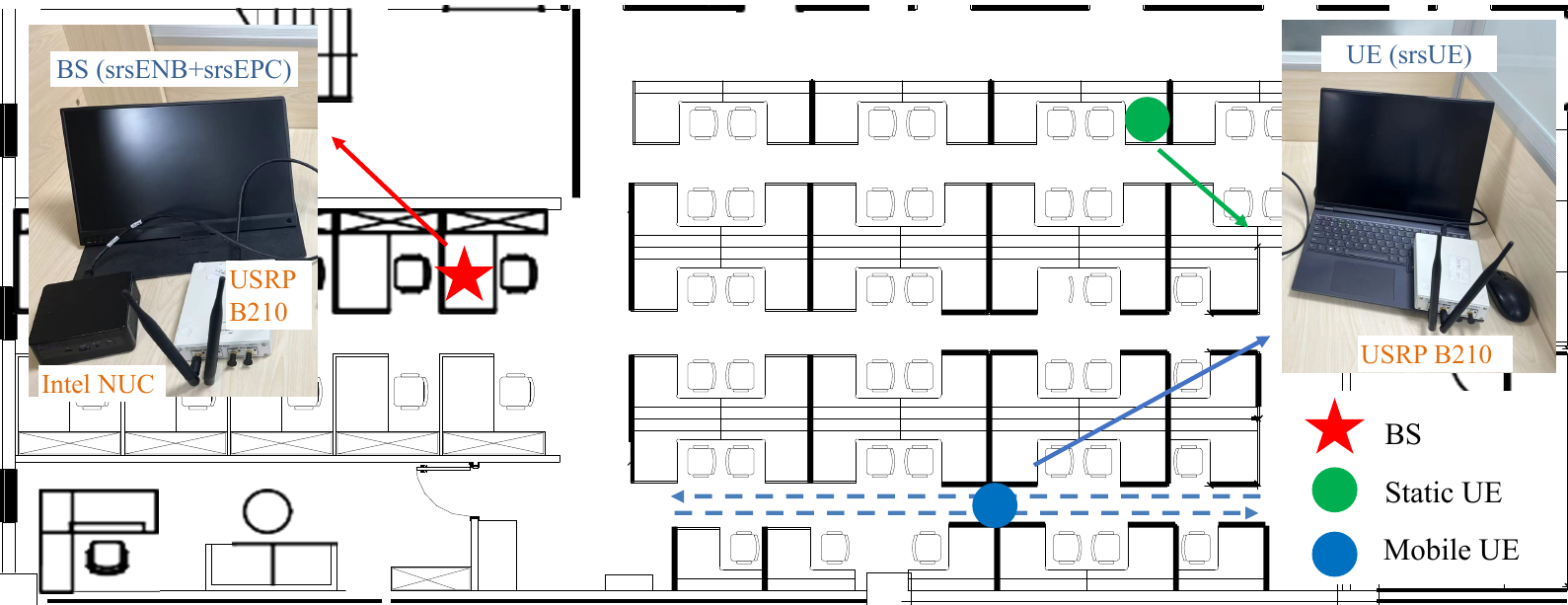}}
\caption{The prototype and the experimental environment where there are a static BS, a static UE and a mobile UE.}
\label{fig equ}
\end{figure}

\subsection{Experimental Setup}

The BS and UE operate in the FDD mode in the 2.68GHz frequency band with a bandwidth of 10MHz. According to the 3GPP LTE standard \cite{3g}, in the FDD system, each frame has a duration of 10ms and comprises 10 subframes. Downlink CSI report is reported from the srsUE side every 40ms, with ACK information feedback available in each TTI. Table \ref{table:system-parameters} summarizes the cellular system parameters and the DC-DQN parameters.

As shown in Fig.~\ref{fig equ}, we conducted experiments in our lab, containing various materials such as machinery, pillars, tables, and walls, which can cause signal reflections and lead to a multi-path channel. We evaluate the performance of algorithms under both the static and mobile scenarios. In the experiments, the BS device is statically positioned in a corner of the laboratory. In the static scenario, we place the UE device at about 8 meters from the BS device, and obstcles exists between the BS device and the UE device. In the mobile scenario, we carry the UE device to walk along a predefined route within the laboratory. The movement introduces changes to the channel condition. By testing our algorithm under both static and moving conditions, we aim to comprehensively evaluate its robustness and adaptability to various link conditions.

We compare the performance of the proposed DC-DQN, \emph{BayesLA}\cite{bayes} and \emph{OLLA}\cite{hsdpa}. We use the default OLLA implementation in srsRAN, and re-implement BayesLA as in \cite{bayes}. In the experiments, we use the \emph{NC} tool to generate the downlink traffic. To saturate the system, we send a large video file with test duration larger than 1000 seconds. Then, we record the logging information (e.g., ACK/NACK) at the BS, and compute the link-layer throughput and block error rate (BLER). In particular, we compute the short-term average throughput over a window of 2000 TTIs.

\subsection{Experimental Results} \label{sec:eval:results}

\begin{figure}[t!] %
\centerline{\includegraphics[width=0.5\textwidth]{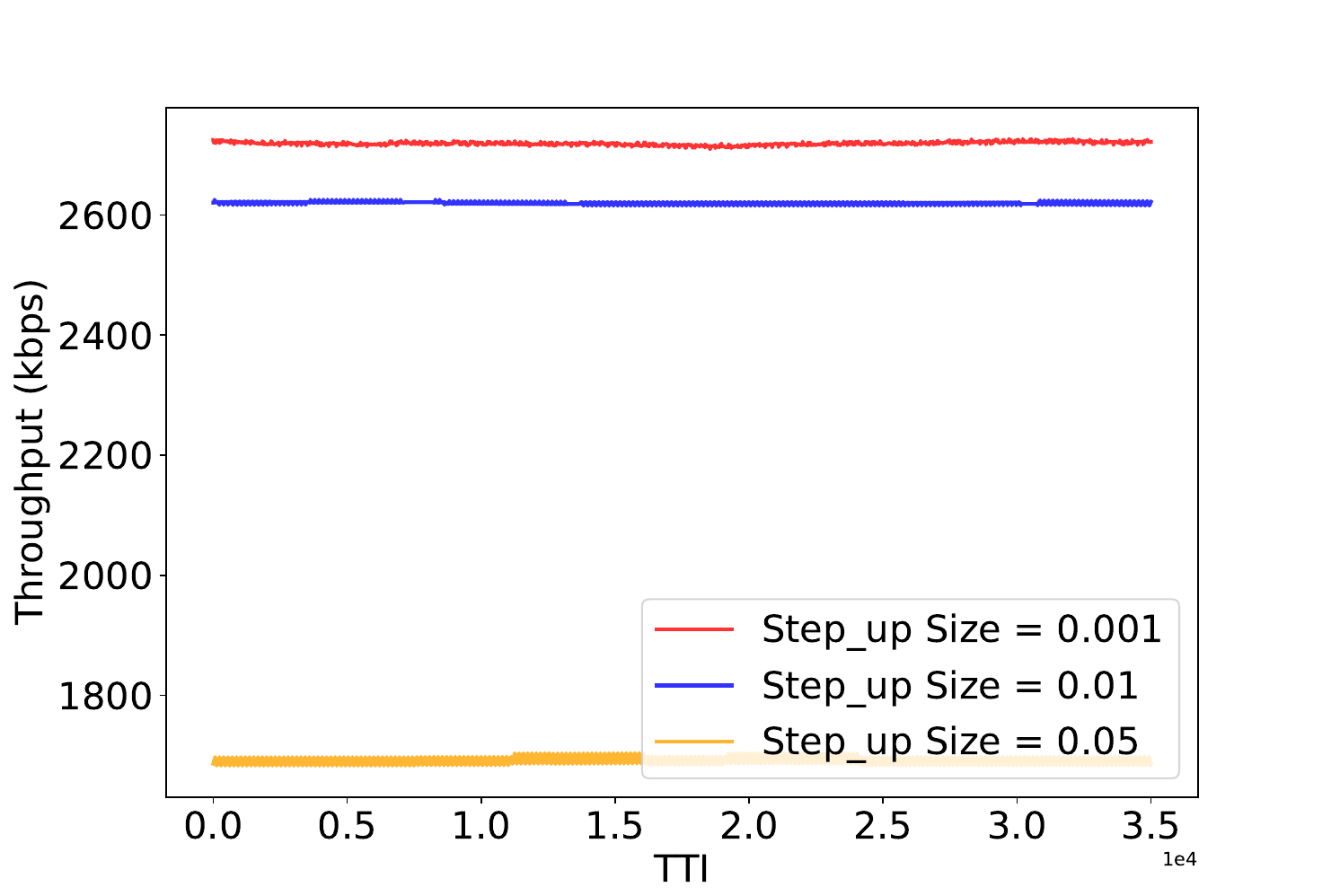}}
\caption{Throughput evolution of various OLLA step sizes.}
\label{OLLA step}
\end{figure}

\textbf{The Static Scenario.}
We first explore the OLLA step size selection in the following experiment and examine the impact of OLLA step size on OLLA performance on the real-world testbed. Fig.~\ref{OLLA step} illustrates OLLA performance across step sizes of $\Delta_{up}$: 0.001, 0.01, and 0.05. Notably, varying step sizes yield significant differences in OLLA's performance within the real-world actual system. In static scenarios, relatively optimal performance is achieved with a step size of $\Delta_{up}=0.001$. As the step size increases, performance degrades. This degradation is attributed to the relatively stable channel conditions. The smaller step size is enough for OLLA to adapt to channel variations and select appropriate MCS based on CQI and offset. Conversely, larger step sizes can cause fluctuation after OLLA convergence, leading to diminished performance. This underscores OLLA's limited adaptability, where optimal step sizes vary across scenarios without automatic adjustment capabilities. In dynamic environments, the lack of adaptability ultimately results in bad performance. In the following experiments, we fixed the step size of OLLA to 0.001.

\begin{figure}[t!] %
\centerline{\includegraphics[width=0.5\textwidth]{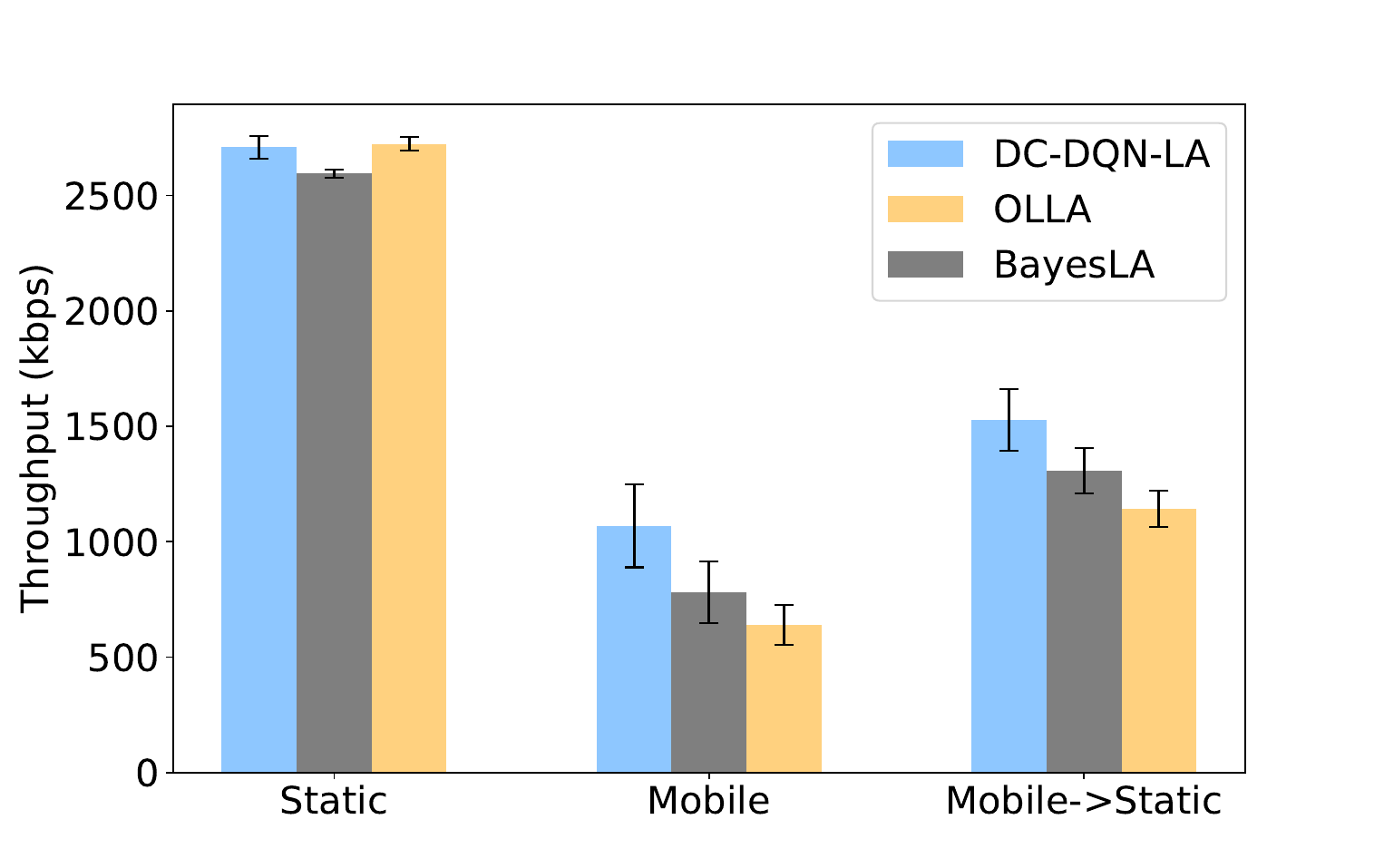}}
\caption{Throughput of various LA techniques in different scenarios on the real-world testbed.}
\label{fig SCA}
\end{figure}

\begin{figure}[t!] %
\centerline{\includegraphics[width=0.5\textwidth]{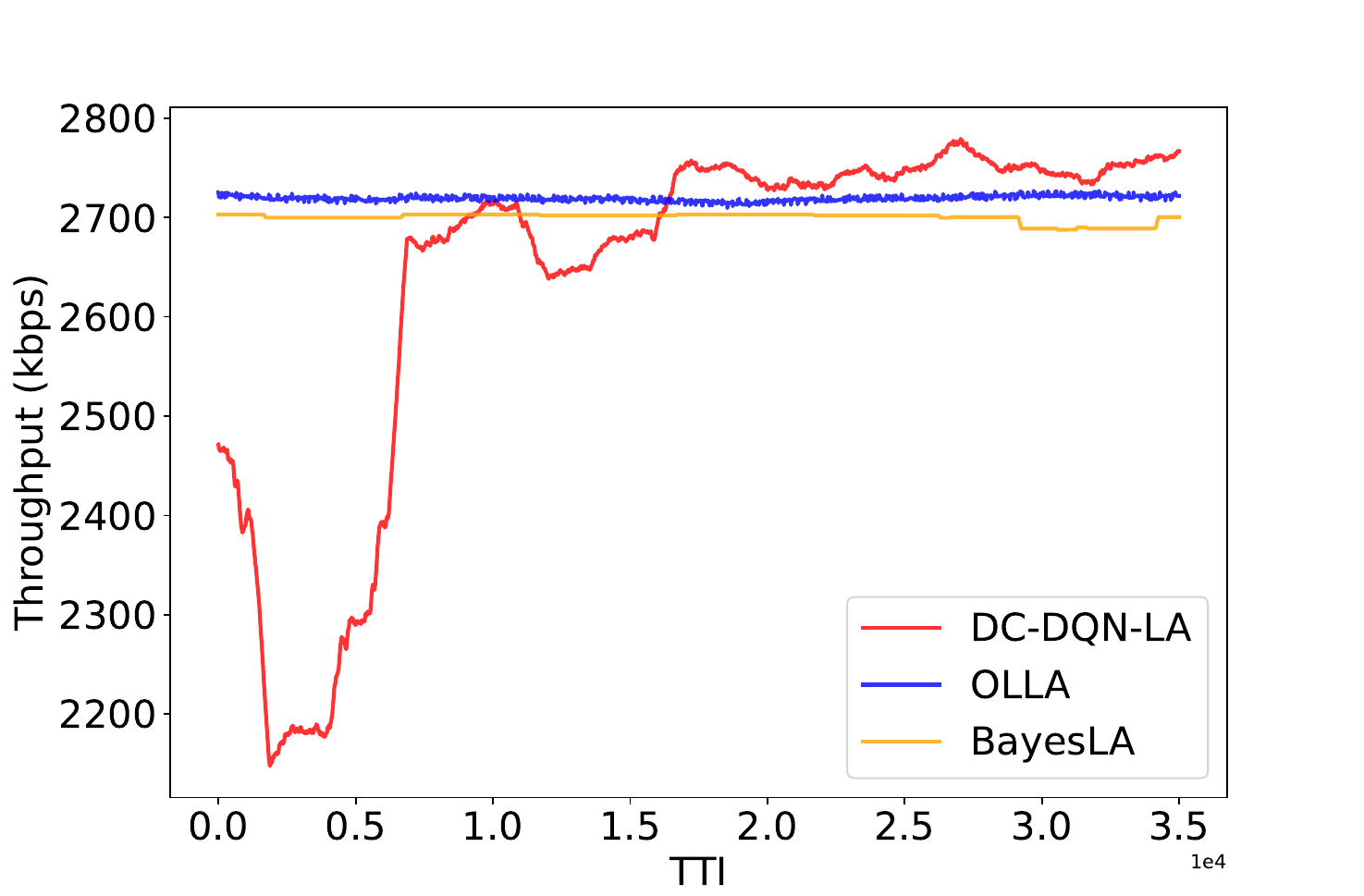}}
\caption{Throughput evolution of various LA techniques in a static scenario on the real-world testbed.}
\label{fig srsran jz}
\end{figure}

The static scenario histogram in Fig.~\ref{fig SCA} shows the throughput of different LA technologies under static channel condition in the real world. We can see that DC-DQN-LA still achieves higher performance, slightly higher than BayesLA and OLLA. Fig.~\ref{fig srsran jz} shows the throughput evolution over time of different LA technologies in the experiment. DC-DQN-LA can converge to a relatively high throughput after a period of training and exploration, although deep reinforcement learning has a certain degree of randomness, but the overall throughput curve remains stable, demonstrating the effectiveness of our algorithm in static scenarios.

\begin{figure}[t!] %
\centerline{\includegraphics[width=0.5\textwidth]{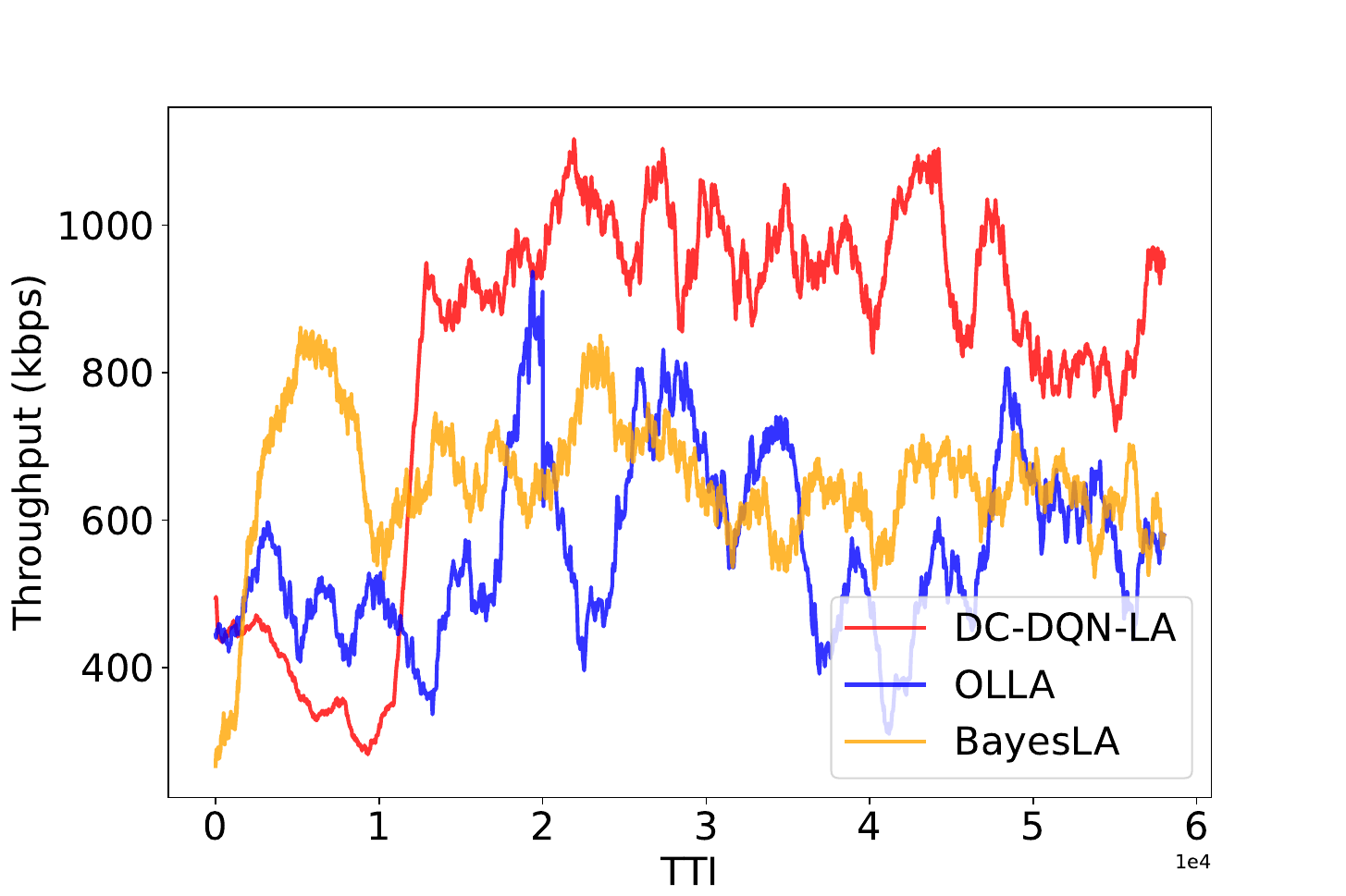}}
\caption{Throughput evolution of various LA techniques in a mobile scenario on the real-world testbed.}
\label{fig srsran yd}
\end{figure}

\textbf{The Mobile Scenario.}
We deploy the above-mentioned algorithm online for training, and evaluated the throughput performance through multiple experiments on the real-world testbed.
Fig.~\ref{fig SCA} presents a histogram illustrating the average and variance of throughput across various LA technologies in multiple real-world experiments under mobile channels. Specifically, Fig.~\ref{fig srsran yd} depicts the evolution of throughput for different LA technologies over time in a specific experiment. It is evident that DC-DQN-LA achieves stable high throughput in actual systems, surpassing BayesLA and OLLA. Specifically, in mobile scenario, DC-DQN-LA achieves a throughput approximately 70\% higher than OLLA and 40\% higher than BayesLA. Meanwhile, the average BLERs of DC-DQN-LA, OLLA and BayesLA are 0.14, 0.1 and 0.23, respectively. DC-DQN-LA slightly increases the BLER compared with OLLA.

Overall, throughput curves of all LA technologies fluctuate significantly over time due to the time-varying and uncertain nature of real channels. Our well-designed decoupling architecture enables strict control over inference time and minimizes computing cost. It is suitable for the practical deployment of DRL-based LA algorithms in communication systems, allowing for the full exploitation of the performance advantages of DRL. %

\begin{figure}[t!] %
\centerline{\includegraphics[width=0.5\textwidth]{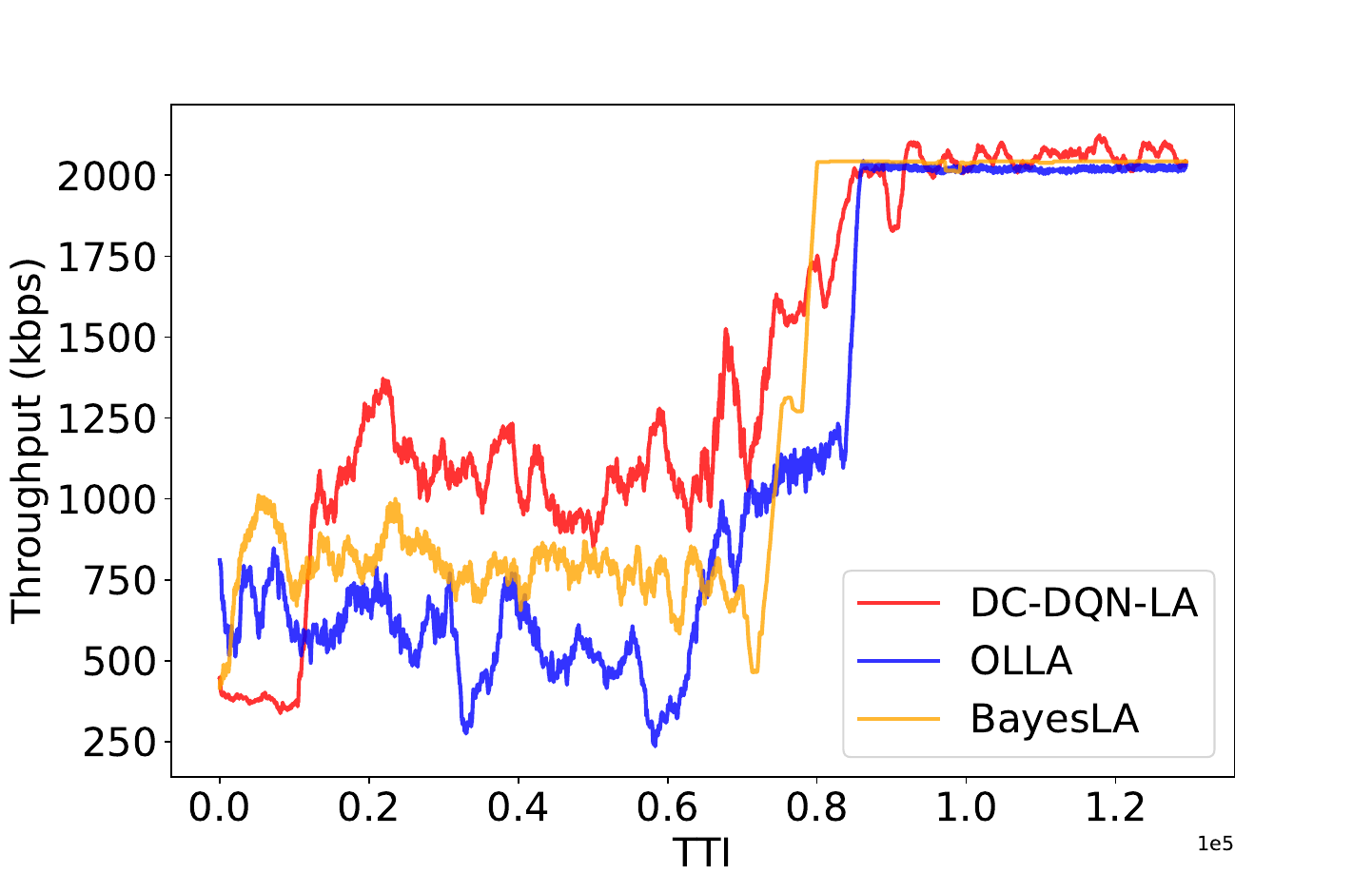}}
\caption{Throughput evolution of different LA techniques in a mobile-to-static scenario on the real-world testbed.}
\label{fig yd jz evo}
\end{figure}

\textbf{The Mobile-to-Static Scenario.}
Next, we evaluated the adaptability of DC-DQN-LA in the real-world changeable scenarios on the testbed. Specifically, at first, the UE continues to move, and the channel is a mobile channel at this time. Then the UE stops moving at TTI 70000, and the channel becomes a static channel. We observe the throughput evolution of DC-DQN-LA under this dynamic change. Fig.~\ref{fig SCA} shows that the overall throughput of DC-DQN-LA is approximately 35\% higher than OLLA and 15\% higher than BayesLA. As shown in Fig.~\ref{fig yd jz evo}, at the moving stage, DC-DQN-LA quickly converges and learns the optimal MCS decisions, achieving significantly higher throughput compared to OLLA approximately 70\%. 

As the UE changes to the static state, DC-DQN-LA gradually adjusts the action and collects experience for parameter tuning. Thus, in subsequent TTIs, DC-DQN-LA relearns suitable MCS decisions to align with the static channel conditions, achieving the stable throughput higher than BayesLA and OLLA. It means that DC-DQN-LA can automatically tracks and adjusts to environment changes without the need for manual parameter reconfiguration. Even employing model parameters trained on prior mobile scenarios, DC-DQN-LA swiftly converges in new static environments, showing its remarkable generalization ability. 

\begin{figure}[t!] %
\centerline{\includegraphics[width=0.5\textwidth]{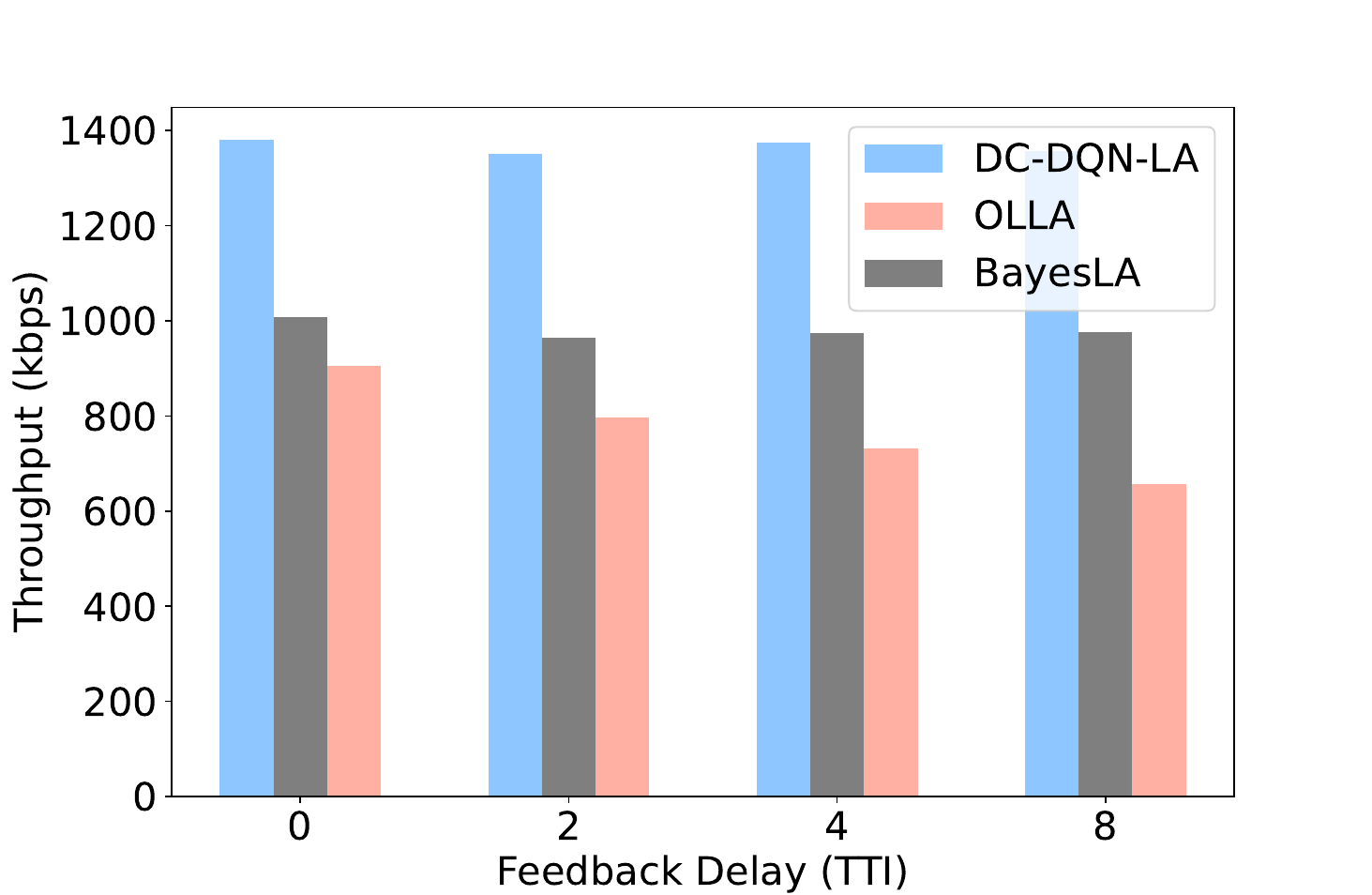}}
\caption{Impact of different feedback delays on the throughput of various algorithms.}
\label{ack_delay}
\end{figure}

\begin{figure}[t!] %
\centerline{\includegraphics[width=0.5\textwidth]{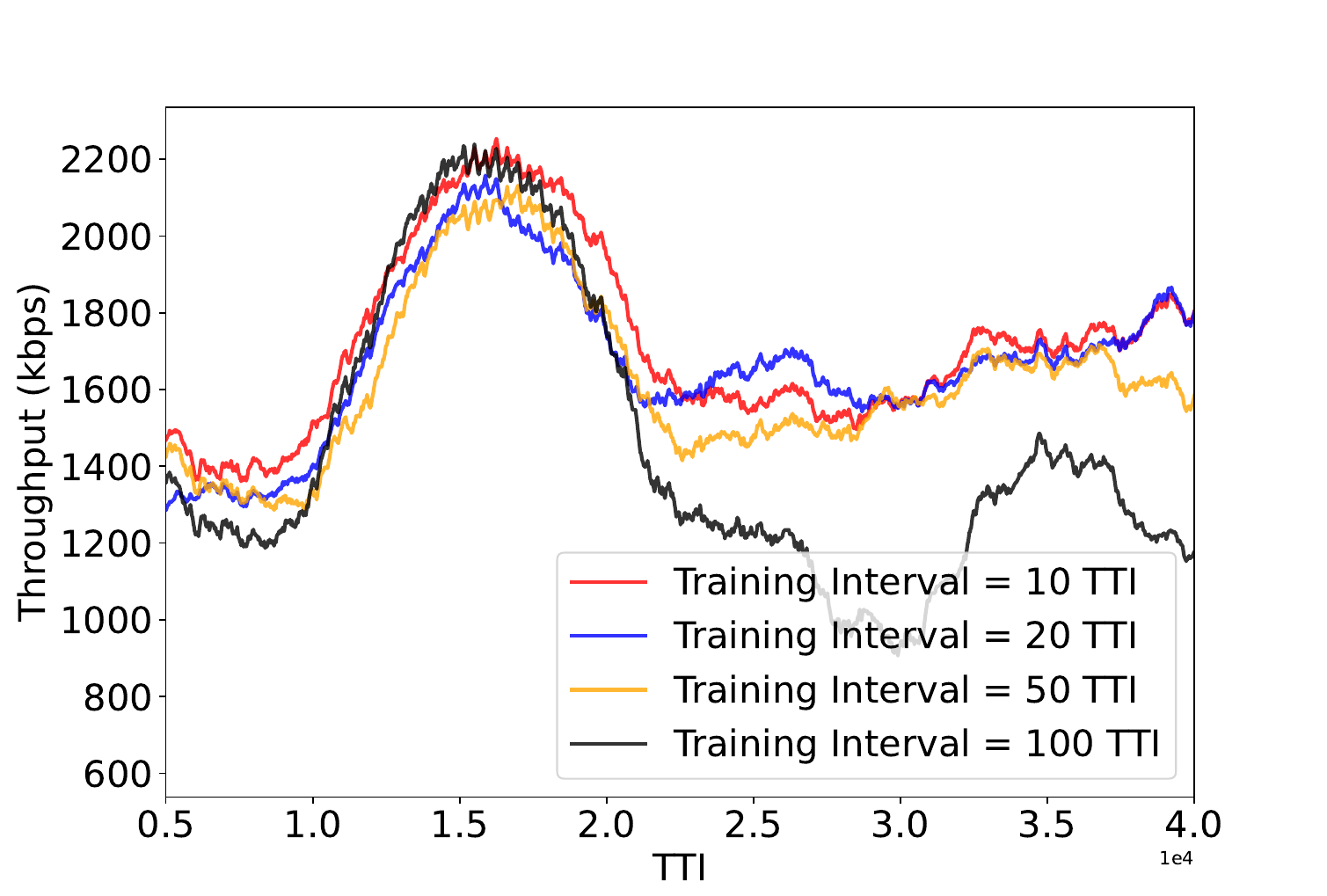}}
\caption{Impact of different training intervals on the throughput of DC-DQN-LA.}
\label{train interval}
\end{figure}

\subsection{Trace-driven Simulation Results}

\subsubsection{Feedback Delay Tolerance} 

In real-world communication systems, there is a fixed delay from making an action to receiving feedback. We investigate the impact of ACK feedback delay on the performance of various algorithms on collected mobile channel via simulations. Specifically, we examine four different delay settings: 0ms, 2ms, 4ms, and 8ms. Fig.~\ref{ack_delay} illustrates the throughput of each method under these varying ACK delay.
The results indicate that DC-DQN-LA consistently outperforms OLLA and BayesLA in terms of link throughput, regardless of the delay situation. Even with the introduction of ACK delay, DC-DQN-LA maintains excellent throughput levels. Specifically, as the delay increases from 0ms to 8ms, DC-DQN-LA's throughput decreases by only 0.3\%, 2.7\%, and 6.9\%, respectively, compared to the throughput without ACK delay. Although there is a slight reduction in performance, DC-DQN-LA still outperforms BayesLA and OLLA by a significant margin.

In contrast, the performance of the OLLA algorithm decrease significantly with the introduction of an 8ms ACK delay, experiencing a throughput decrease of approximately 30\%. This decline is attributed to delayed feedback cause adverse effects on the adjustment of OLLA's offset, leading to recurrent failures. DC-DQN-LA, on the other hand, adopts a more conservative strategy, which enables it to achieve greater throughput even in the presence of delayed feedback. As the ACK delay increases from 0 to 8 ms, the throughput gain of DC-DQN-LA gradually rises, ranging from approximately 50\% to about 100\% compared to OLLA. It demonstrates the robustness of DC-DQN-LA to different ACK feedback delays, highlighting its ability to maintain consistent high performance in the practical system.

\subsubsection{Training Interval Insensitivity}

The training module of our DC-DQN-LA algorithm can employ flexible training intervals, which significantly reduce computing resource cost by decreasing the frequency of training. To investigate the impact of different training intervals in DC-DQN-LA on performance and select the optimal interval for practical application, we conduct experiments with varying training intervals and measured throughput on the collected mobile channel. The experimental results are depicted in Fig.~\ref{train interval}.

The training interval appears to have minimal impact on the convergence speed of the algorithm. Whether set at 10ms, 20ms, 50ms, or 100ms, DC-DQN-LA generally achieves convergence within approximately 12000 TTIs. Furthermore, the throughput is basically same after the algorithm converges, except when the training interval is set to 100ms, where a significant performance drop occurs. This phenomenon can be attributed to the longer training interval's inability to promptly adapt to changes in channel conditions, resulting in delayed adjustments to the model's behavior. Nevertheless, when the training intervals are 10ms, 20ms, and 50ms, the algorithm achieves similar throughput levels, suggesting that reasonable adjustments to the training interval do not significantly impact performance. This observation underscores the rationality of our algorithm's decoupled structure. Moreover, the algorithm's insensitivity to the training interval allows for deployment of the training module at a low operating frequency without compromising model performance, thereby reducing overhead compared to traditional DRL-based algorithms, proving more suitable for practical systems.

\section{Related Work} \label{sec:related}
\para{LA for Wireless Networks.} OLLA algorithms \cite{hsdpa,degrees} were studied dated back to the 3G cellular networks. Several heuristic algorithms \cite{2016eolla,2015dynamic,2015self} have been developed to improve OLLA. However, these algorithms exhibit limited adaptability in various and unseen link conditions. 
Learning algorithms that directly select MCS through DRL \cite{2019drl,ye2023tmc,parsa2022joint} or multi-armed bandits \cite{saxena2019contextual} are proposed to adapt various environments. Some learning algorithms directly tune the OLLA parameters \cite{saxena2021reinforcement,pang2024efficient}.
We follow the DRL approach that directly selects MCS, and address the practical issues when applying DRLLA to practical cellular networks.

Refs. \cite{karmakar2017smartla, chen2021experience, geiser2022drlla, queiros2022wi, lin2022deep, yin2024adr} propose DRL-based rate adaptation algorithm for WiFi networks, and demonstrate superior throughput compared with the traditional rate adaptation algorithms. %
Similar to these papers, we apply DRL-based link adaptation to improve the throughput of cellular networks. However, we address different challenges, since centralized cellular networks are different from distributed WiFi networks.

\para{DRL Algorithms for Networking Problems.} 
DRL-based algorithms have been used in several networking problems, such as congestion control \cite{jay2019deep}, packet scheduling \cite{chen2018auto} and mobile video telephony \cite{zhang2020onrl}, and have demonstrates superior performance than existing algorithms. 
However, the implementation efficiency remains an issue.

Our work is inspired by the lightweight DRL-based congestion control algorithms \cite{zhang2022liteflow, tian2022spine} in data center networks. They also leverage the two-level control framework but target low resource consumption. Different from the congestion control problem, LA in cellular networks are challenged by the more dynamic environment (e.g., the change time of wireless channel is on the order of ms), the more stringent response time (e.g., TTI is on the order of ms) and the fixed feedback delay defined by the protocol, leading to different designs. %

Refs. \cite{restuccia2020deepwierl, cho2019fa3c} address the DRL efficiency problem from the hardware perspective. In particular, they leverage the highly-parallel FPGA hardware to accelerate DRL implementation. Instead, we focus on the architecture level, and design a DC-DQN-LA framework to improve the efficiency.

\section{Conclusion}

We have put forth a new DRL-based LA algorithm named DC-DQN-LA for practical cellular networks.  DC-DQN-LA avoids the performance degradation problem induced by the excessive computation latency problem of existing DRLLA algorithms with a novel framework design that separates the inference module and the time-consuming training module. Additionally, DC-DQN-LA accounts for the feedback delay and HARQ issues with the new MDP design and the experience alignment technique. Prototype implementation and experimental results show the efficiency and efficacy of the new practical DRLLA algorithm.

\appendices

\section{DC-DQN Convergence Analysis}

To analyze the convergence properties of DC-DQN, we utilize the following theorem from \cite{fan2020theoretical}:

\begin{theorem}\label{def1}
	Let $\epsilon$ be the sampling distribution over $\Omega_{\leq T} \times \mathcal{A}$, $\mu$ be a fixed probability distribution over $\Omega_{\leq T} \times \mathcal{A}$, and $R_{\max}$ be the maximum reward value. Then we have
	\begin{equation}\label{eq converger 2}
		\mathbb{E}_\mu\left[\left|Q^* - Q^{\pi_L}\right|\right] \leq \frac{2\phi_{\mu,\sigma}\gamma}{(1-\gamma)^2} \cdot \eta_{\max,T} + \frac{4\gamma^{L+1}}{(1-\gamma)^2} \cdot R_{\max},
	\end{equation}
	where $\phi_{\mu,\sigma}$ is the concentration coefficient of $\mu$ and $\sigma$, and $\eta_{\max,T}$ is the maximum one-step approximation error given by $\eta_{\max,T} = \max_{t \in T} \|\mathcal{B}Q_{t-1} - Q_t\|$, where $\mathcal{B}(\cdot)$ denotes the Bellman optimality operator.
\end{theorem}

In DC-DQN, both the target network $\theta^t$ and the decision network $\theta^d$ are periodically synchronized with the main network $\theta$ at the same interval $U$:
\textbf{Every $U$ steps, the main network's parameters $\theta$ are copied to both the target network $\theta^t$ and the decision network $\theta^d$ \cite{mnih2015human}.}
This synchronization ensures that both networks remain aligned with the main network, reducing discrepancies that could arise from asynchronous updates. Consequently, decisions are made using stable parameters, enhancing the interaction stability between the network and the environment.

To quantify the difference between the main network $\theta$ and the decision network $\theta^d$, we introduce a delay metric based on the function space:
\begin{equation}\label{delay metric}
	\Delta_t^Q = \max_{s,a} \left| Q_{\theta^d}(s,a) - Q_{\theta}(s,a) \right|.
\end{equation}

This metric measures the maximum difference in Q-values between the decision network and the main network across all state-action pairs at time $t$, capturing the extent to which the decision network lags behind due to periodic updates.

The delay metric $\Delta_t^Q$ impacts both the sampling distribution and the maximum one-step approximation error $\eta_{\max,T}$:
\begin{itemize}
	\item \textbf{Sampling Distribution}: The actions selected by the decision network $\theta^d$ determine the state-action distribution $\sigma_{\theta^d}$ from which experiences are sampled. The difference $\Delta_t^Q$ affects how much $\sigma_{\theta^d}$ deviates from the main network's distribution $\sigma$.
	\item \textbf{Maximum One-Step Approximation Error}: The delay $\Delta_t^Q$ introduces an additional error in the Q-function approximation, affecting $\eta_{\max,T}$.
\end{itemize}

The delay metric $\Delta_t^Q$ influences the sampling distribution $\sigma_{\theta^d}$. To ensure that the sampling distribution does not deviate significantly from the intended distribution $\sigma$, we establish the following relationship:
\begin{equation}
	d(\sigma_{\theta^d}, \sigma) \leq \rho \cdot \Delta_t^Q, 
\end{equation}
where $d(\cdot,\cdot)$ denotes a suitable distance metric (e.g., total variation distance) between distributions, and $\rho$ is a constant that depends on the Lipschitz continuity of the policy with respect to the Q-values \cite{precup2001off}.

To control $\Delta_t^Q$, we adjust the learning rate $\alpha$ and the update period $U$:
\begin{itemize}
	\item \textbf{Learning Rate $\alpha$}: By choosing a sufficiently small learning rate, we limit the step size of updates, thereby controlling how quickly $\theta$ changes and, consequently, how much $\theta^d$ can differ from $\theta$ over $U$ steps.
	\item \textbf{Update Period $U$}: A smaller $U$ means more frequent updates of $\theta^d$, reducing the window during which $\theta$ can diverge from $\theta^d$, thereby keeping $\Delta_t^Q$ smaller.
\end{itemize}

Given these controls, we can bound $\Delta_t^Q$ as:
\begin{equation}
	\Delta_t^Q \leq L \cdot U \cdot \alpha,
\end{equation}
where $L$ is the Lipschitz constant of the Q-function.

The maximum one-step approximation error $\eta_{\max,T}$ is thus influenced by the delay $\Delta_t^Q$ as:
\begin{equation}
	\eta_{\max,T} \leq \eta_0 + \Gamma \cdot \Delta_t^Q,
\end{equation}
where $\eta_0$ is the base approximation error without delay, and $\Gamma$ is a constant that captures how the delay $\Delta_t^Q$ amplifies the approximation error, depending on the properties of the Q-function and the Bellman operator.

With the introduction of delay, the sampling distribution becomes $\sigma_{\theta^d}$, and the concentration coefficient is updated accordingly:
\begin{equation}
	\phi_{\mu,\sigma_{\theta^d}} = \sup_{h \in \mathcal{H}} \frac{\mathbb{E}_{\mu}[h(s,a)]}{\mathbb{E}_{\sigma_{\theta^d}}[h(s,a)]}.
\end{equation}

Given that $\sigma_{\theta^d}$ is influenced by $\Delta_t^Q$, we can bound the change in $\phi_{\mu,\sigma}$ as:
\begin{equation}
	\phi_{\mu,\sigma_{\theta^d}} \leq \phi_{\mu,\sigma} + \Psi \cdot \Delta_t^Q,
\end{equation}
where $\Psi$ is a constant that depends on the sensitivity of the distribution conversion to changes in the Q-values \cite{jin2020provably}.

Incorporating delay into Theorem 1, the bound becomes:
\begin{equation}
	\begin{split}
		\mathbb{E}_{\mu}\bigl[|Q^{*} - Q^{\pi_L}|\bigr] &\leq \frac{2 (\phi_{\mu,\sigma} + \Psi \cdot \Delta_t^Q) \gamma}{(1 - \gamma)^2} \cdot (\eta_0 + \Gamma \cdot \Delta_t^Q) \\
		&\quad + \frac{4 \gamma^{L+1}}{(1 - \gamma)^2} R_{\max}.
	\end{split}
\end{equation}
The higher-order terms are neglected due to the smallness of $\Delta_t^Q$.

To formally establish that DC-DQN with delay maintains the convergence properties of traditional DQN, we proceed under the following assumptions:
\begin{itemize}
	\item The Q-function satisfies Lipschitz continuity with constant $L$:
	\begin{equation}
		\|Q_{\theta}(s,a) - Q_{\theta'}(s,a)\| \leq L \|\theta - \theta'\|
	\end{equation}
	
	\item Concentration coefficient bound:
	\begin{equation}
		\phi_{\mu,\sigma_{\theta^d}} \leq \phi_{\mu,\sigma} + \Psi \cdot \Delta_{\max}
	\end{equation}
\end{itemize}

Under these conditions:
\begin{itemize}
	\item \textbf{Control of $\Delta_t^Q$}: To control the delay metric $\Delta_t^Q$, we appropriately set the update interval $U$ and the learning rate $\alpha$, ensuring that $\Delta_t^Q \leq \Delta_{\max}$, where $\Delta_{\max}$ is a predefined threshold.

	\item \textbf{Bounded Distribution Difference}: This constraint guarantees that the distribution conversion coefficient satisfies $\phi_{\mu,\sigma_{\theta^d}} \leq \phi_{\mu,\sigma} + \Psi \cdot \Delta_{\max}$, where $\Psi$ is a constant that depends on the sensitivity of the distribution conversion to changes in the Q-values \cite{mnih2015human}.

	\item \textbf{Bounded Approximation Error}: Substituting the bounded $\Delta_t^Q$ into the modified Theorem 1. Since $\Delta_{\max}$ is small, the additional terms involving $\Psi \cdot \Delta_{\max}$ and $\Gamma \cdot \Delta_{\max}$ are negligible, effectively maintaining the original convergence bound with minor adjustments.
	
	\item \textbf{Maintaining Convergence}: The adjusted bound retains the form of the original Theorem 1, with an additional term accounting for the controlled delay. Therefore, as long as $\Delta_t^Q$ is sufficiently small, DC-DQN maintains the convergence properties of traditional DQN.
\end{itemize}

By ensuring that $\Delta_t^Q$ is kept small through appropriate choices of learning rate $\alpha$ and update period $U$, the additional error introduced by the delay remains negligible. Consequently, the convergence properties of DC-DQN are preserved and remain consistent with those of traditional DQN.

\ifCLASSOPTIONcaptionsoff
  \newpage
\fi

\bibliographystyle{IEEEtran}
\bibliography{ref}

\end{document}